\documentclass[preprint,11pt,eqsecnum]{aastex}

\slugcomment{ApJ in press}

\shorttitle{Thermal Physics of X-winds}
\shortauthors{Shang et al.}

\begin{document}
\newcommand{\sech}{\rm sech}
\newcommand{\xe}{x_{\rm e}}
\newcommand{\xh}{x({\rm H})}
\newcommand{\mh}{{\rm H}_2}
\newcommand{\xmh}{x({\rm H}_2)}
\newcommand{\xhm}{x({\rm H}^-)}
\newcommand{\hm}{{\rm H}^-}
\newcommand{\hp}{{\rm H}^+}
\newcommand{\Lst}{L_{\ast}}
\newcommand{\Tst}{T_{\ast}}
\newcommand{\Rst}{R_{\ast}}
\newcommand{\Wst}{W_{\ast}}
\newcommand{\gAst}{g_{{\rm A}, \ast}}
\newcommand{\TAst}{T_{{\rm A}, \ast}}
\newcommand{\gS}{g_{\rm S}}
\newcommand{\LX}{L_{\rm X}}
\newcommand{\TX}{T_{\rm X}}
\newcommand{\tauX}{\tau_{\rm X}}
\newcommand{\Lh}{L_{\rm h}}
\newcommand{\Th}{T_{\rm h}}
\newcommand{\Fh}{F_{\rm h}}
\newcommand{\Wh}{W_{\rm h}}
\newcommand{\gAh}{g_{\rm A,h}}
\newcommand{\TAh}{T_{\rm A,h}}
\newcommand{\TS}{T_{\rm S}}
\newcommand{\kS}{k_{\rm S}}
\newcommand{\Rx}{R_{\rm x}}
\newcommand{\Ox}{\Omega_{\rm x}}
\newcommand{\zetax}{\zeta_{\rm x}}
\newcommand{\Mw}{M_{\rm w}}
\newcommand{\vw}{v_{\rm w}}
\renewcommand{\baselinestretch}{1.35}
\newcommand{\vdag}{(v)^\dagger}
\newcommand{\ds}{\displaystyle}
\newcommand{\be}{\begin{equation}}
\newcommand{\ee}{\end{equation}}
\newcommand{\tbar}{{\langle t \rangle}}
\newcommand{\Rmax}{R_{\rm max}}
\newcommand{\Rmin}{R_{\rm min}}
\newcommand{\rhoi}{\rho_{\rm i}}
\newcommand{\rhon}{\rho_{\rm n}}
\newcommand{\fL}{{\bf f}_{\rm L}}
\def\eg{{e.g.\ }}
\def\etc{{etc.\ }}
\def\etal{\mbox{\it et al.\,}}
\def\ie{{i.e.\ }}

\title{Heating and Ionization of X-Winds}
\author{Hsien Shang\altaffilmark{1,2}, Alfred E. Glassgold\altaffilmark{3}, 
Frank H. Shu\altaffilmark{2}} 
\affil{Astronomy Department, University of California, Berkeley, CA 94720}
\email{hshang@cfa.harvard.edu,fshu,aeg@astro.berkeley.edu}
\and
\author{Susana Lizano}
\affil{Instituto de Astronom\'\i a, UNAM, 58090 Morelia, Michoac\'an, M\'exico}
\email{lizano@astrosmo.unam.mx}

\altaffiltext{1}{Present address: Center for Astrophysics, 60 Garden Street,
Cambridge, MA 02138}
\altaffiltext{2}{Institute of Astronomy and Astrophysics, Academia Sinica,
Taipei, TAIWAN}
\altaffiltext{3}{Physics Department, New York University, NY 10003}
\begin{abstract}
In order to compare the x-wind with observations, one needs to be able to 
calculate its thermal and ionization properties.  We formulate the physical 
basis for the streamline-by-streamline integration of the ionization and
heat equations of the steady x-wind.  In addition to the well-known
processes associated with the interaction of stellar and
accretion-funnel hot-spot radiation with the wind, we include X-ray
heating and ionization, mechanical heating, and a revised calculation of
ambipolar diffusion heating. The mechanical heating arises from
fluctuations produced by star-disk interactions of the time dependent
x-wind that are carried by the wind to large distances where they are
dissipated in shocks, MHD waves, and turbulent cascades. We model the
time-averaged heating by the scale-free volumetric heating rate, 
$\Gamma_{\rm mech} = \alpha \rho v^3 s^{-1}$, where $\rho$ and $v$ are
the local mass density and wind speed, respectively, $s$ is the
distance from the origin, and $\alpha$ is a phenomenological
constant. When we consider a partially-revealed but active young stellar object,
we find that choosing $\alpha \sim 10^{-3}$ in our numerical
calculations produces temperatures and electron fractions that are
high enough for the x-wind jet to radiate in the optical forbidden
lines at the level and on the spatial scales that are observed. We
also discuss a variety of applications of our thermal-chemical
calculations that can lead to further observational checks of x-wind 
theory.
\end{abstract}

\hfill
\newpage

\section{Introduction}

	A refined and updated version of the disk-accretion paradigm for
the formation of Sun-like stars has emerged in the last two decades 
through extensive observations and theoretical studies. In
addition to the building up of the new star by accretion from a
disk formed by the collapse of a rotating molecular cloud
core, the formation of a low-mass star is accompanied by a remarkable
bipolar outflow that can appear jet-like at optical and nearby
wavelengths.  Equally important is the crucial role of magnetic fields
in retarding the initial collapse and in guiding both the accretion
flow that feeds the star and the outflow that removes excess angular
momentum. In addition to the strong evidence provided by the essentially
universal detection of X-rays in low-mass young stellar objects
(YSOs), magnetic fields have been measured directly with the Zeeman
effect (e.g., Johns-Krull and Valenti, 2000). Although the general
outline of a theory of low-mass star formation has emerged, the
underlying mechanisms still need to be identified and understood, and strong
efforts along these lines are in progress on a broad front, as can be 
seen in the reports at the recent conference {\it Protostars and
Planets IV} (Mannings, Boss, and Russell 2000).

	One of the main goals of the theory is to develop a
rational description of the active flows close to the central engine,
i.e., the accretion funnel and the wind. Although there is a consensus
that these flows are MHD in character, considerable disagreement
exists over the specifics, as witnessed by the reviews of
magnetocentrifugal winds at {\it Protostars and Planets IV} by
K\"onigl and Pudritz (2000; disk winds) and by Shu, Najita, Shang, and
Li (2000; x-wind), as well as the earlier review of wind theory by
Pudritz and Ruden (1993). Of course the only way to decide between
alternative theories or to validate any particular theory is to make
detailed comparisons with observations.  Thus, it is the objective of the
present paper to develop the basis for making such comparisons for the
case of the x-wind (Shu \etal 1994a; Shu \etal 1994b; Najita \& Shu
994; Ostriker \& Shu 1995; Shu \etal 1995; henceforth Papers I-V).
The x-wind model has the potential for understanding
many aspects of low-mass star formation because it provides
well-defined dynamical solutions that can be used to make detailed
correlations and predictions of observational data.  It should be
clear that, because x-wind theory focuses on the inner region (or
``central engine'') of the star in formation of dimension 0.1\,AU, 
 observational tests require spectroscopy on the milli-
arc-second spatial scale and better. Such observations are now
becoming available with adaptive optics and interferometric techniques, 
as well as with the Hubble Space Telescope.

        As discussed by Shu \etal (2000), several crucial assumptions
and implications of the x-wind model are supported by observations:
the existence of a finite inner radius for a rapidly rotating inner
disk; strong magnetization of the central star; magnetically channeled
accretion; and phase relations between stellar rotation and the
accretion funnel and the outflow. The model can also account for the
large-scale kinematic properties of bipolar molecular outflows (Shu
\etal 1991, 2000), and it has the potential to explain the optical
observations of jets from young stellar objects. The latter
possibility is a consequence of the remarkable property of the x-wind,
whose 180$^{\circ}$ bipolar lobes self-collimate towards the outflow
axis into approximately cylindrical jets (Paper V).  On this basis,
Shang, Shu, \& Glassgold (1998, henceforth SSG) made synthetic images
of optical jets that bear a striking resemblance to the observed ones
(see, e.g., Eisl\"offel \etal 2000).

        These conclusions have been made on the basis of the dynamical
solution of the x-wind model which gives the density, velocity, and
magnetic field configuration\footnote{The main physical requirements for the
solution is that the electron fraction be large enough to justify the
MHD approximation and the temperature be low enough to ignore thermal
pressure.}. The images obtained by SSG were obtained by using {\it
constant} values of the temperature and electron fraction based on
previous analyses of the optical observations. The main goal of the
present paper is to lay the foundation for obtaining the physical
properties of the x-wind from first principles in order to calculate
the distribution of the temperature, the ionization fraction, and
other chemical abundances for the x-wind streamlines. These properties
are required for calculating the fluxes of diagnostic lines and
continua that observers can use to test the validity of the model.

        Even though the dynamics of the x-wind are largely decoupled
from the thermal-chemical properties, the calculation of these
properties for a 2-dimensional flow is very difficult.  The closest
previous work by Ruden, Glassgold, \& Shu (1990, henceforth RGS; see
also Glassgold, Mamon, \& Huggins 1991 for a parallel chemical study) 
done before the x-wind solutions were obtained, assumed spherical
symmetry. They tried to anticipate some aspects of the 2-d
axisymmetric solutions by modulating the radial density and velocity
variations at small distances $r$. RGS found that radial winds quickly
cool and become weakly ionized with increasing $r$. The importance of
the run of ionization and temperature in outflows is well illustrated
by the observations of optical jets, where phenomenological analyses
of line strengths indicate that they are significantly ionized and
hot, with electron fractions $x_{\rm e} \sim 0.01-0.1$ and
temperatures $T \sim 5,000 - 10,000$\,K (e.g., Bacciotti 2001).
Shocks have long been the favored mechanism for producing these
conditions (e.g., Raga, B\"ohm, and Cant\`o 1996; Hartigan, Bally, Reipurth,
and Morse 2000). Aside from the ``final'' bow shocks with the ambient medium,
associated with the Herbig-Haro objects, it has been unclear how the central
young stellar object (YSO) can affect the global properties of the jet at large
distances from the source.  Bacciotti \etal (1995) suggested that the jet
retains a high level of ionization characteristic of the source region by
virtue of slow radiative recombination in the flow. Our calculations support
this idea, once we add an important missing ingredient, the ionization of
the base of the wind by the X-rays that are observed to accompany
essentially all YSOs (e.g., Feigelson \& Montmerle 1999).

        Not only are the X-rays effective in ionizing the wind close
to the source, they also help maintain the ionization level at large
distances. The emissivity of the flow is determined by the
temperature, and it is essential for understanding jets to also be
able to achieve the high temperatures indicated by the observations of
the optical forbidden lines.  Here shocks can play an important role,
and we will develop a global model of wind heating based on stochastic
shock dissipation. Although our model of mechanical heating is
supported by MHD simulations (e.g., Ostriker et al. 1999), at this
stage it is essentially phenomenological. By adopting physically
reasonable parameters, we will be able to obtain images of jets that
resemble the observations. We should also be able to relieve the
difficulties encountered by RGS, since the wide-angle component of the
x-wind seems likely to be warm and moderately ionized. In a
reconsideration of ambipolar diffusion heating, we will find that a
new atomic coefficient provides a much reduced role for this process
in warm atomic regions, in disagreement with earlier results by Safier
(1993).

        The main goal of this paper is to develop a coherent
thermal-chemical foundation for the x-wind. We build on the previous
study by RGS, but add important new processes for ionization (X-rays)
and heating (mechanical or turbulent shock heating). As the main
illustrative application, we consider the jets as observed in optical
forbidden lines. The rest of this paper is organized as follows.  In
the next section \S 2, we give a general description of the model, and
in the following sections we focus on X-ray ionization (\S 3),
ambipolar diffusion heating (\S 4), and mechanical heating (\S5).
Modeling results are then presented in \S 6 for the case of an active
solar-mass YSO in a partially revealed phase. We then discuss further
implications of the calculations in \S 7 in the context of future
detailed studies that bear directly on observations. The paper
contains a number of appendices that supplement the technical basis
for \S\S 2-5.

\section{Formulation of the Model}

	The calculation of the thermochemical properties of the x-wind
is separable from the dynamical problem (treated in Papers I-V) in the
cold ideal MHD limit.  As long as the electron fraction is large
enough and the thermal energy is small enough (for the thermal
pressure to be small compared with the kinetic and magnetic energies),
the MHD approximation can be made.  We of course check that the
thermochemical calculations reported here satisfy these
assumptions. Most of this section will be devoted to formulating the
thermal and chemical equations that need to be solved along each
streamline. As discussed in \S 2.1, the streamlines are obtained on
the basis of the dynamical solution for the x-wind obtained by Shang
(1998).

\subsection{Dynamics}

	The exact numerical solution for the x-wind in Paper III does
not provide a practical basis for thermochemical modeling because it
is restricted to the sub-Alfv\`enic region.  Similarly, the asymptotic
solution in Paper V does not apply at small distances.  We use a 
semi-analytic approach developed by Shang (1998),
where a global x-wind solution is obtained by interpolating between
the two extreme solutions developed in Paper II and Paper V.  This 
solution applies to the steady and axisymmetric X-wind flow that we
model here.  It satisfies the conservation laws of mass, specific
angular momentum, and energy, which are expressed in terms of the
conserved quantities $\beta(\psi)$, $J(\psi)$, and $H(\psi)$ on the
streamlines $\psi$ as shown in Paper II.  In steady state, the field
lines co-rotate with the star at angular velocity $\Omega_\ast$.
Unlike Papers II and III, which employ a reference frame that rotates
with the stellar angular velocity $\Omega_\ast$, we describe the flow
in an inertial frame.  We follow Paper I in using non-dimensionalized
equations based on the units for length, velocity, density, and
magnetic field: $\Rx,\;\Ox\Rx,\;\dot{M}_w/4\pi \Rx^3\Ox,
\;(\Ox\dot{M}_w/\Rx)^{1/2}$, where $\Ox=\Omega_\ast$ is the angular
velocity at the X-point $\Rx$, and $\dot{M}_w$ is the mass-loss rate of
the x-wind. Our calculation is restricted to the case treated in 
Papers I-V of aligned magnetic and rotation axes. Many of the same 
physical processes discussed in the following sections would be 
operative in the more general case, but the required dynamical solutions 
do not exist yet.

In the asymptotic regime, an x-wind streamline $\psi$ is represented in 
spherical coordinates $r-\theta$ by the parametric equations,
\begin{equation}
\label{asymp1}
r_a = \frac{2\bar{\beta}}{C}\cosh[F(C,1)]
\end{equation}
and 
\begin{equation}
\label{asymp2}
\theta_a=\sin^{-1} \{\sech [F(C,\psi)]\},
\end{equation} 
where 
\begin{equation}
F(C,\psi)\equiv \frac{1}{C} \int_0^{\psi}\frac{\beta (\psi')d\psi'}
{[2J(\psi')-3-2C\beta (\psi')]^{1/2}}, 
\end{equation}
and the subscript $a$ stands for asymptotic.  The stream function
$\psi$ has the range $0 \leq \psi \leq 1$ and labels the mass fraction
carried by the x-wind from the horizontal plane.  The quantity $\beta$
is the ratio of the magnetic-to-mass-flux, i.e., MHD field-freezing in
the co-rotating frame implies that the magnetic field and mass flux
are proportional to one another: ${\bf B}=\beta\rho{\bf u}$.  The
condition $\nabla \cdot {\bf B} = 0$ requires that $\beta$ be
conserved on streamlines, i.e., $\beta = \beta(\psi)$.  Similarly, the
conservation of total specific angular momentum requires that the
amount carried by matter in the inertial frame, $\varpi v_\varphi$,
plus the amount carried by Maxwell torques, $-\varpi B_\varphi {\bf
B}$, per unit mass flux, $\rho {\bf u}$, is a function of $\psi$
alone: $\varpi [v_\varphi -\beta^2\rho (v_\varphi-\varpi)] = J(\psi)$.
The function $C$ is a slowly decreasing function of $r$ introduced in
Paper V that vanishes logarithmically when $r \to\infty$, consistent
with the vanishing of the current at infinity, independent of
streamline.  The locus $\psi=1$ is obtained by applying approximate
pressure balance, $B_z^2=B_\varphi^2$, across the wind-deadzone
interface.

In the cold limit, the function $\beta(\psi)$ cannot be chosen
completely arbitrarily; otherwise the magnetic field, mass flux, and
mass density will diverge on the uppermost streamline ($\psi
\rightarrow 1$) as the x-wind leaves the x-region.  For modeling
purposes, Shang (1998) adopts the following distribution of magnetic
field to mass flux:
\begin{equation}
\label{bt0}
\beta (\psi) = \beta_0 (1-\psi )^{-1/3},
\end{equation}
where $\beta_0$ is a numerical constant related to the mean value of $\beta$
averaged over streamlines: 
\begin{equation}
\label{barbeta}
\bar\beta \equiv \int_0^1 \beta (\psi) \, d\psi = {3\over 2}\beta_0.
\end{equation}
The singularity in equation (\ref{bt0}) is of no real concern: it 
reflects the fact that the magnetic field ${\bf B} = \beta \rho {\bf
u}$ is nonzero on the uppermost x-wind streamline, where by definition
$\rho$ must become vanishingly small while ${\bf u}$ remains finite.

In order to make contact with the inner solution, we follow 
Paper II and use pseudopolar coordinates $(s, \varphi, \vartheta)$ 
with the origin of $s$ at the x-point and the angle $\vartheta$ measured 
in the meridional plane starting from zero at the equator:
\be
\varpi=1+s\cos\vartheta,\hspace{0.25in} z=s\sin\vartheta.
\ee
When $r \gg 1$, $s\approx r$ and $\vartheta\rightarrow \pi/2-\theta$.
Equations~(\ref{asymp1}) and~(\ref{asymp2}) now become, 
\label{asymp}
\begin{eqnarray}
{s_a}=\frac{2\bar{\beta}}{C}\cosh[F(C,1)], \\
{\vartheta_a}=\cos^{-1} \{ \sech [F(C,\psi)]\}.
\end{eqnarray}
Near the x-point, the magnetic field lines emerge uniformly and define 
a fan within a $60^{\circ}$ sector above the equatorial plane 
($0 \leq \vartheta_f \leq \pi/3$), 
\be
\label{fan}
{\vartheta_f (\psi)=\vartheta_f(0) \frac{1}{\bar{\beta}}
\int_0^{\psi} \beta({\psi'})d\psi'},
\ee
where the subscript $f$ stands for fan. The uppermost streamline with 
$\psi =1$ emerges at $\vartheta_f = \pi/3$ and the lower-most streamline with 
$\psi =0$ emerges at $\vartheta_f = 0$.

The interpolation between the outer asymptotic and inner fan solutions 
is based on the equations,
\be
\label{interp1}
s(C)=\frac{2\bar{\beta}}{C}\cosh[F(C,1)]
-\frac{2\bar{\beta}}{C_x}\cosh[F({C_x},1)], 
\ee
and
\be
\label{interp2}
\vartheta(C,\psi)= \alpha\vartheta_f(\psi)
+(1-\alpha){\vartheta_a}(C,\psi).
\ee
The second term in equation~(\ref{interp1}) (with the constant $C_x$) 
has been introduced so that the function $s=s(C)$ reduces to zero at 
the x-point ($s(C_x)=0$).  The reference value $C_x$ is chosen so that
equation~(\ref{asymp2}) yields $\pi/3$,
\ie, the asymptotic streamlines occupy the same angular region as the fan
emerging from the x-point (and are in approximate pressure balance). 
An interpolation function $\alpha(C,\psi)$ was chosen to make a smooth 
transition between the fan and the asymptotic solutions.

\subsection{Heating and Ionization}

        The equations governing the temperature $T$ and electron fraction 
$x_{\rm e}$ are
\be
\label{heateq}
{3\over 2}{DT\over Dt} = 
T{D \over Dt} \ln n(1+x_{\rm e} + x_{\rm He})^{-3/2} + 
{1\over 1+x_{\rm e} + x_{\rm He}}\left({\cal G} - {\cal L}\right), 
\ee
\be
\label{ioneq}
{Dx_{\rm e}\over Dt} = {\cal P}-{\cal D} .
\ee
where $D/Dt = {\bf v}\cdot \nabla$ is the substantial time derivative
along a streamline, ${\cal P} \equiv P/n$ and ${\cal D} \equiv D/n$
are ionization production and destruction rates (dimensions s$^{-1}$),
and ${\cal G} \equiv \Gamma /nk$ and ${\cal L} \equiv \Lambda/nk$ are
heating and cooling rates (dimensions K\,s$^{-1}$).  The Latin symbols
$P$, $D$, $G$, and $L$ are the corresponding volumetric rates, and $n$
 is the total number of hydrogen nuclei per unit volume (usually
denoted $n_{\rm H}$ in the literature). We use $n$ to define
abundances, e.g., $x_{\rm e}= n_{\rm e}/n$ is the electron fraction,
$x_{\rm He}$ is the He abundance, and $(1+x_{\rm e} + x_{\rm He})n$
is the total number density of particles ignoring elements heavier
than He.  Many of the source terms in equations~(\ref{heateq}) and
(\ref{ioneq}) were formulated in a unified way by RGS, and we use their
expressions for individual terms in $P$, $D$, $G$, and $L$ whenever
possible. We discuss those terms in RGS which require change in \S 2.4
and appendix A and B, and new ionization and heating sources in \S\S
2-5. Table~\ref{tbl-1} lists the processes included in the present study.

\placetable{tbl-1}
\begin{deluxetable}{ll}
\tabletypesize{\footnotesize}
\tablecolumns{2}  
\tablewidth{0pc}  
\tablecaption{Physical Processes\label{tbl-1}}
\tablehead{\multicolumn{2}{c}{Production of Ionization}}
\startdata
H$^{-}$ photodetachment                 & \S 2.4	     \\
Balmer continuum photoionization        & \S 2.4, App.~B     \\      
Electronic collisional ionization       & \S 2.4             \\
X-Rays                                  & \S 3, App.~C 	     \\
\cutinhead{Destruction of Ionization}
Radiative recombination                 & \S 2.4       \\      
\cutinhead{Heating}
Photodetachment of H$^{-}$              & \S 2.4             \\
Balmer continuum photoionization of H   & \S 2.4             \\      
H$^{+}$-H$^{-}$ neutralization          & \S 2.4, App.~A     \\
Ambipolar Diffusion                     & \S 4, App.~D, App.~E  \\
X-Rays                                  & \S 3          \\
Mechanical 	                        & \S 5          \\
\cutinhead{Cooling}
Adiabatic                               & \S 2.2        \\      
H$^-$ radiative attachment              & \S 2.4           \\
Recombination of H$^+$                  & \S 2.4          \\
Lyman $\alpha$                          & \S 2.4          \\
Collisional ionization                  & \S 2.4        \\
Heavy element line radiation            & \S 2.5        \\
\enddata
\end{deluxetable}

Two basic time scales are those for adiabatic cooling and
recombination. Near the source, their values for the fiducial model
introduced in \S 6 are of the order of tens of seconds and days,
respectively.  We usually omit explicit consideration of the effects
of the ionization of He, and adopt $x_{\rm He}=0.1$ in calculations described
in this paper.  In the regions of main concern, the central part of the wind in
and around the jet, the dominant ionizing agents are the secondary electrons
produced following X-ray ionization of heavy elements.  As long as $x_{\rm e} >
3 \times 10^{-3}$, the inclusion of He would increase $x_{\rm e}$ by the factor
$(1+ x_{\rm He})^{1/2}$ or 5\%.

\subsection{The Accretion Hot Spot}

The x-wind may be heated and ionized by several sources of radiation.
In addition to the photosphere, these include a ``hot spot'' produced
by the infall of the accretion funnel onto the stellar surface; the
accretion disk, which scatters and re-emits absorbed stellar radiation
and radiates energy generated by viscous dissipation; and the
star-disk magnetosphere, which emits thermal and nonthermal radiation
in both soft ($< 1$\,keV) and hard ($> 1$\,keV) X-rays. Because we are
mainly concerned with the upper streamlines (close to the axis) that
constitute the bulk of the mass of the inner wind, we can safely
ignore the disk radiation. Not only is the mean photon energy smaller
than for stellar radiation, the radiation emitted by the disk is
significantly diluted by geometry and absorbed by the wind. The
effects of disk radiation and gas-dynamic interactions with the disk
atmosphere are important for the lower streamlines, which constitute
the wide angle wind. We focus here on the hot spot radiation and treat
the X-rays in detail in \S 3.

We model the hot spot following Ostriker \& Shu (1995).
For an axisymmetric x-wind, the hot spot is an annulus located between 
co-latitudes $\theta_1$ and $\theta_2$. The hot spot luminosity is
\be
\label{hotspotlum}
\Lh = \left (1-f \right )\frac{G M_\ast \dot{M}_{\rm D}}{R_\ast}
\left[1+{R_\ast^3 \sin^2\bar{\theta}_{\rm h}\over 
2R_{\rm x}^3}-{3R_\ast\over 2R_{\rm x}} \right],
\ee
where $f$ is the fraction of the disk accretion rate $\dot{M}_{\rm D}$
that is lost by the x-wind (and thus $(1-f)$ is the mass transfer rate of
the funnel flow). The hot spot covers a fraction $F_{\rm h} = \cos
\theta_1-\cos \theta_2$ of the surface area of the star, and the mean
colatitude of the hot spot is $\sin \bar{\theta}_{\rm h} = F_{\rm h}/
( \theta_2- \theta_1)$.  We use parameters for the fiducial case 
in \S 6 that are appropriate for the preferred model of 
Ostriker \& Shu: $\theta_1=26.6^{\circ}$ and $\theta_2= 33.2^{\circ}$, 
and with $R_\ast/R_{\rm x} = 0.20$, $F_{\rm h} \approx 0.06$ and $\sin^2
\bar{\theta}_{\rm h} \approx 0.5$. Calvet \& Gullbring (1998) and
Gullbring \etal (2000) have analyzed the spectral energy distributions
of T Tauri stars with an accretion shock model based on a dipolar
magnetic field.  They find that lightly veiled T Tauri stars have
$F_{\rm h}$ in the range $0.001-0.05$.  Such small hot spots can be
realized with a generalization of x-wind theory that brings in higher
multipole fields (Mohanty \& Shu 2001).

The presence of the hot spot is a potential complication in the 
radiation transfer needed to calculate the absorption of radiation by 
the wind. We could achieve some simplification by 
exploiting the axial symmetry of the steady x-wind and replacing the 
hot spot by equivalent sources located at the origin and along the polar 
axis of symmetry. Because we are mainly interested in radial distances 
$r >> R_\ast$, however, we can simply replace the hot spot by a source 
located at the origin and calculate the mean intensity of the 
hot spot radiation as a blackbody of temperature $\Th$ with the
approximate dilution factor,
\be
\label{hotspotdil}
\Wh = \Fh {R_\ast^2\over 4r^2}.
\ee
The effective temperature $\Th$ is given by
\be
\label{hotspottemp}
\Th^4 = T_\ast^4 + \left(\frac {\Lh}{\Fh 4\pi R_\ast^2\sigma}\right),
\ee
if we assume that the accretion shock structure is optically thin to
the underlying stellar (photospheric) continuum and denote the
Stefan-Boltzmann constant by $\sigma_{\rm B}$.  The corresponding
dilution factor for the stellar photospheric radiation is
\be
\label{stardil}
W_\ast= (1-\Fh) {R_\ast^2\over 4r^2}.
\ee
The mean intensities of the stellar and the hot spot radiation fields
before they enter the wind are obtained by multiplying the dilutions 
factors by $\sigma_{\rm B}\Th^4$ and  $\sigma_{\rm B}\Tst ^4$, 
respectively. Although neither the stellar nor the hot-spot radiation
is accurately represented by a blackbody spectrum at
wavelengths shortward of the Balmer continuum, the blackbody approximation does
not lead to any serious error in this paper because the X-rays 
are the dominant ionization source (\S 3).

\subsection{Radiative and Collisional Processes for Hydrogen}

The stellar and hot spot radiation fields are important in the inner
wind for detaching the electron of the negative hydrogen ion (in the
continuum shortward of $ \lambda = 1.647\,\mu$m) and for photoionizing
the $n=2$ level of atomic hydrogen (in the Balmer continuum shortward
of $ \lambda = 0.365\,\mu$m).  These processes and their inverses were
discussed in detail by RGS for a stellar blackbody, and we describe
only the most important changes in their methodology. RGS expressed
the rates for these photoionization processes in the form,
\be
\label{RGSgenericP} 
{\cal P} = W \, g_{\rm A}x, 
\ee 
where $x$ is the abundance of the species (H$^-$ or H(2), where we write
H(n) for the H atom with principal quantum number n), $W$ is the
standard dilution factor for a star, and $g_{\rm A}$ is the 
absorption rate for the appropriate continuum. Applying the results 
of \S 2.3, our ionization rates are,  
\be
\label{genericP}
{\cal P_{\rm A}} = \left (\Wst \, \gAst + \Wh \, \gAh \right ) x,
\ee  
where $\Wst$ and $\Wh$ were defined in \S 2.3 and $\gAst$ and $\gAh$ 
are absorption rates evaluated using blackbody spectra at the stellar 
and hot spot temperatures, respectively.  
         
\subsubsection{The Negative Hydrogen Ion}

The rate equation for $\hm$ is:
\be
\label{H-rateeq}
{D\over Dt}\xhm = \xe (1-\xe)\kS(H^-)n
- \left[ \Wh g_{A,h}(H^-)+\Wst g_{A,\ast}(H^-)+x_e n k_\pm\right]\xhm.
\ee
In addition to photodetachment, $\hm$ is also destroyed rapidly by
neutralization with $\hp$ in which a hydrogen atom is produced in an
excited state, usually $n=3$ for energies less than 1 eV (e.g., Fussen
\& Kubach 1986). The inverse reaction, H(1) + H(3) $\rightarrow \hm +
\hp$, contributes little to the production of $\hm$, mainly because
the abundance of excited H(3) atoms is so small. The potentially
important reaction, H(1) + H(1) $\rightarrow \hm + \hp$, is
endothermic by 12.85 eV and has a relatively small cross section even
well above threshold. It can therefore be safely ignored as a
production mechanism for $\hm$. We use a rate coefficient $k_\pm$
somewhat larger than in RGS, as explained in Appendix A.  We calculate
the photo rates in equation~(\ref{H-rateeq}) by numerically integrating
the Wishart (1979) photodetachment cross section $\alpha_\nu(\hm)$ in
RGS Eq.~(A16),
\be
\label{gS}
\gS(\hm,T) \equiv 4\pi \int_{\nu_-}^\infty 
\frac{\alpha_\nu(\hm)}{h\nu}
\left (\frac{2h\nu^3}{c^2}\right ) e^{-h\nu/kT}d\nu, 
\ee
and in RGS Eq.~(A18),
\be
\label{gAhm}
\gAst(\hm,\Tst)\equiv 4\pi\int_{\nu_-}^\infty\frac{\alpha_\nu(\hm)}
{h\nu}B_\nu(\Tst)d\nu,
\hspace{0.25in}
\gAh(\hm,\Th)\equiv 4\pi\int_{\nu_-}^\infty\frac{\alpha_\nu(\hm)}
{h\nu}B_\nu(\Th)d\nu,
\ee
where $\nu_-$ is the threshold frequency for $\hm$ photodetachment 
at $\lambda_- = 1.647\,\mu$m.  
The subscript S in equation~(\ref{gS}) labels $\gS$ as a spontaneous rate. We use 
detailed balance (RGS Eq.~[20]) to obtain the rate coefficient for 
photodetachment from $\gS$,
\be
\label{H-detailbal}
\kS(\hm,T)\equiv \frac{\lambda^3_e}{4}e^{h\nu_0/kT}\gS(T).
\ee
We note that $\gAst(\hm,\Tst)$ and $\gAh(\hm,\Th) $ are functions of the
stellar and hot spot temperatures, whereas $\gS(\hm,T)$ is a function of
kinetic temperature. We need to calculate $\gAst(\hm,\Tst)$ and
$\gAh(\hm,\Th)$ just once at the beginning of a calculation, but we need
to know $\kS(\hm,T)$ everywhere on each streamline.  In order to save
computing time, we have fit $\kS$ by a formula similar to that
introduced by Stancil, Lepp, \& Dalgarno (1998),
\be
\label{kS}
\qquad \kS(\hm,T) = 
1.33\times 10^{-18}\,(T^{0.85}+T^{0.4})(e^{-T/9320}+e^{-T/18000}),
\ee
which reproduces the numerical calculations to better than 15\% over
the temperature range from $10^2 - 2\times 10^4$\,K. We have ignored 
attenuation in equation~(\ref{gAhm}) because the optical depth in the $\hm$ 
continuum is small due to the low abundance of $\hm$. 

Associated with the gain and loss terms in the rate equation for $\hm$, 
equation~(\ref{H-rateeq}), are the heating and cooling rates, 
\be
{\cal G}(\hm) = \left [\Wh  \gAh(\hm,\Th)   \TAh(\hm) + 
                       \Wst \gAst(\hm,\Tst) \TAst(\hm) \right ]\xhm, 
\ee 
\be
\label{hminuscool}
{\cal L}(\hm) = \xe(1-\xe) n \kS(\hm,T)\TS(\hm,T).
\ee
The heating rates (in K units), $\gAh(\hm,\Th) \TAh(\hm)$ and
$\gAst(\hm,\Tst) \TAst(\hm)$, are given by integrals like those in
equation~(\ref{gAhm}) except that an additional factor $(\nu-\nu_-)$
appears in the integrands, as is appropriate to photoelectron
heating. The cooling rate $\kS(\hm,T)\TS(\hm,T)$ is given by a similar
modification of the equation for $\gS$. Again we evaluate $\TAh$ and
$\TAst$ numerically once per calculational case and use the following
fit for $\TS(\hm,T)$
\be
\label{TShm}
\TS(\hm,T)= 
T\left(\frac{\xi_-^2+4\xi_-+6}{\xi_-^2+2\xi_-+2}\right) 
\left(e^{-T/50000}+ e^{-T/4000}\right ),
\ee
where $\xi_- = T_-/T = 8750\,{\rm K}/T$; this approximation is accurate to 
within a few percent over the temperature range from $10^2 -10^5$\,K.  

\subsubsection{The $n=2$ Level of Atomic Hydrogen}

Our treatment of the H atom is essentially standard Case B
recombination theory modified for a 2-d axisymmetric wind by use of
the Sobolev approximation (e.g., Shu 1991, Chapter 9). It goes 
beyond RGS by including electronic collisional and X-ray ionization 
and excitation. The broader implications of the X-rays are 
taken up in \S 3. We adopt a simplified two-level 
plus continuum model for the H atom, and formulate and solve the 
steady-state population equations in Appendix B. The population of
the $n=2$ level is given in the convention defined after 
equation~(\ref{heateq}) by, 
\be
\label{n=2pop}
x_2 = \frac{(1-\xe)}{1+Q}, \qquad 
Q = \frac{k_{21}n \xe + A_{21}\beta_{21}}
              {k_{12}n \xe + k_{1c} n \xe + \zeta_{1c} + \zeta_{12}},
\ee
where $A_{21}$ is the spontaneous radiative decay rate of the 
Lyman-$\alpha$ line, $\beta_{21}$ is the corresponding escape probability, 
\be
\beta_{21} = {1-e^{-\tau_{21}}\over \tau_{21}}, 
\ee
and $\tau_{21}$ is the (locally-calculated) optical depth of the line,  
\be
\label{lymanalphaod}
\tau_{21} = {\pi e^2\over m_e c}\lambda_{21} f_{12}(1-x_e)nS^{-1} , \qquad
S = {2\over 3}{v_w\over s} + {1\over 3}{dv_w\over ds}.
\ee
The rate coefficients $k_{12}$ and $k_{21}$ describe electronic
collisional excitation and de-excitation of the $n=2$ level; $k_{1c}$ is
the rate coefficient for electronic collisional ionization of the $n=1$
level. The quantities $\zeta_{1c}$ and $\zeta_{12}$ are respectively 
the rates at which X-rays ionize and excite the
H-atom ground state, as described in \S 3 and Appendix~B. 
The main additional changes from RGS are that the radial
coordinate $r$ is replaced by the streamline coordinate $s$ and the
characteristic length $2v_w / 3r$ is replaced by $S$ in 
equation~(\ref{lymanalphaod}).

Our treatment of photoionization by the Balmer continua and the
recombination into levels $n \geq2$ closely parallels the above
formulation for $\hm$. It is actually simpler because the integrals 
over frequency can be done in closed form when the cross section 
has the Kramers $\nu^3$ dependence (RGS Appendix
B). The coefficients in the  photoionization rates for the stellar 
and hot spot blackbodies in equation~(\ref{genericP}) are then 
given by integrals like those in equation~(\ref{gAhm}), with 
$\alpha_\nu(\hm)$ replaced by the Kramers cross section (RGS Eq.~[21]); 
the result is
\be
\label{gA2}
g_{A,h}(2) = e^{-\tau_{2c}}
{8\pi\over c^2}2\nu_2^3 \alpha_1 \sum_{j=1}^{\infty}
E_1(jT_{2}/T_{\rm h}), \qquad
g_{A,\ast}(2) = e^{-\tau_{2c}}
{8\pi\over c^2}2\nu_2^3 \alpha_1 \sum_{j=1}^{\infty}
E_1(jT_{2}/T_{\ast}),   
\ee
where $\nu_2$ is the frequency threshold for photoionization from the
$n=2$ level and $T_{2}$ is the same quantity in temperature units.
The factor $\exp(-\tau_{2c})$ reflects the fact that we include
absorption of the Balmer continuum, which is well calculated by
replacing the frequency dependent attenuation factor
$\exp(-\tau_{\nu})$ by its threshold value. The optical depth at
threshold is
\be
\label{Balmerdepth}
\tau_{2c} = 2\alpha_1 N_2,
\ee
where 
\be
\label{n=2column}
N_2 = \int_0^r x_2 n \,dr,
\ee 
is the radial column density for the $n=2$ level. 

The rate at which radiative recombination occurs is given by,
\be
\label{recombrate}
{\cal D}_e = \xe^2 n\kS(\ge 2).
\ee
We calculate the rate coefficient $\kS(\ge 2)$ from RGS (their Eq.~[B15]
and preceding equations). As discussed by RGS, the evaluation involves
replacing sums over the principal quantum
number $n$ from $n=2$ to $n_{\rm max}$ by integrals.  RGS determined $n_{\rm max}$ from
the condition derived by Seaton (1964) for the equality of the rates
for collisional ionization and spontaneous decay for large quantum
numbers. The Seaton condition characterizes an important property
of the large $n$ level population (where the rate of increase of the
departure coefficients begins to decrease), but it does not describe
the size of the atom, i.e., $n_{\rm max}$. In fact, for the conditions
in the wind, $n_{\rm max}$ is determined by Stark broadening (e.g. Mihalas 1970),
\be
\label{nmax}
n_{\rm max} = 1234\, (n_{\rm e}\, {\rm cm}^3)^{-2/15}.
\ee
Despite the small exponent, $n_{\rm max}$ ranges from $\approx 30$
at the base of the wind to $100-200$ at large distances. This
variation has little effect on the value of the recombination rate
coefficient, but it does have a bearing on the observability of sub-mm
radio recombination lines (\S 6).

The photoionization of the $n=2$ level and the recombination to levels $n\geq 2$ 
lead to heating and cooling, respectively:
\be
{\cal G}(2) = \left [\Wh \gAh(2) \TAh(2)  + \Wst \gAst(2) \TAst(2) \right ]x_2, 
\ee 
\be
\label{recombcool}
{\cal L}(\geq 2) = \xe^2 n \kS(\geq 2)\TS(\geq 2).
\ee The heating energies, $\TAst(2)$ and $\TAh(2)$, are calculated
from the closed form expression, RGS Eq.~(B8), and the recombination
cooling $\kS(\geq 2)\TS(\geq 2)$ from RGS Eqs.~(27), (B13), and (B16).

The heating and cooling by the H atom occurs by collisional excitation,
de-excitation, and ionization and by photoionization and electronic
recombination.  When the net effect of collisional excitation and 
de-excitation are expressed in terms of the standard formula for 
Lyman-$\alpha$ line cooling,
\be
\label{lymanalphacooling}
{\cal L}_{\rm Ly-\alpha} =  x_2A_{21}\beta_{21}T_{21}.
\ee
the heating is given by (Eq.~[\ref{finalheateq}]),
\be
{\cal G} - {\cal L} = -{\cal L}_{\rm Ly-\alpha} + {\cal G}(2)
-{\cal L}(\geq 2) - {\cal L}_{\rm coll} + {\cal G}^{\prime}_{\rm X},
\ee
where 
\be
\label{collioncooling}
{\cal L}_{\rm coll}  = \frac{1}{4}\xe (k_{1c}n_1 + k_{2c}n_2)T_{1c}  
= \frac{1}{4}\xe(1-\xe)nT_{1c}
\left  ( \frac{k_{1c}}{1 + Q^{-1}} + \frac{k_{2c}}{1+ Q} \right ) ,  
\ee
and
\be
\label{indirectX}
{\cal G}^{\prime}_{\rm X} = 2.22 \zeta T_{12} ;
\ee
$T_{1c}$ and $T_{12}$ are the ionization potential and the excitation
energy of the first excited level of atomic hydrogen (in K units), and
$k_{1c}$ and $k_{2c}$ are the rate coefficients for collisional
ionization of the $n=1$ and $n=2$ levels. As discussed in Appendix B,
equation~(\ref{indirectX}) defines an indirect X-ray heating term that
arises because we use equation~(\ref{lymanalphacooling}) to eliminate
the net heating from collisional heating and cooling in favor of
Lyman-$\alpha$ cooling. This term arises from the effects of the
X-rays on the population of the $n=2$ level of atomic hydrogen. The
collisional ionization rates in equation~(\ref{collioncooling}) are
discussed in Appendix B.

\subsection{Heavy Element Cooling}

For the temperatures important in the inner wind, the heavy elements
mainly contribute to the cooling by forbidden line transitions, many
of which serve as observational diagnostics for the jets that emanate
from YSOs.  We have included the forbidden transitions of O\,I, S\,II,
and N\,I at solar abundances using 5-level model ions as in SSG. N\,I cooling
is about one order of magnitude smaller than O\,I and S\,II cooling in our
calculations. We have not included Fe\,II which, by virtue of a large number of
transitions, can be an important coolant at high densities. Only in the last
few years have realistic calculations of Fe cooling been published (Woitke and
Sedlmayr 1999; Verner \etal 1999, 2000). According to the cooling functions
displayed graphically by Woitke and Sedlmayr, Fe\,II cooling should be small
compared to the dominant adiabatic cooling in our wind models. The reason for
this is that, for the densities characteristic of the x-wind jet, the strength
of the Fe\,II cooling is determined by the abundance of iron compared to
oxygen. It is only at much higher densities, where the Fe\,II lines become
optically thick, that the very large number of Fe\,II lines makes this ion a
more powerful coolant than more abundant ions with simpler level structure. 
 
\section{X-Rays}

The ionization and heating of the x-wind by X-rays are another example
of how X-rays produced by YSOs affect the physical conditions for star
formation. Earlier we used observations of YSO X-rays to determine the
flux of energetic particles that can produce short-lived radionuclides
at the time the Sun was formed (Shu \etal 1997; Lee \etal 1998;
Gounelle \etal 2001). We have also found that the observed level of
X-rays is more than sufficient to provide the dominant source of
ionization in the atmospheres of protoplanetary disks (Glassgold,
Najita, \& Igea 1997, henceforth GNI; Igea \& Glassgold 1999). Here we
show how these X-rays affect the degree of ionization at the base of
the jets observed in low-mass stars.

	The ROSAT measurements of nearby clusters of newly formed
stars provide us with a good idea of the average soft X-ray flux
from T Tauri stars, and ASCA has extended these results to higher
X-ray energies albeit at lower angular resolution (e.g., Feigelson \&
Montmerle 1999). The mean soft X-ray luminosity for a typical T-Tauri
star is $\LX \sim 5 \times 10^{29}$\,erg s$^{-1}$. The satellite
observatories have also begun to elucidate the dependence of the X-ray
emission on the age of the YSO, the amount of variability, and the
properties of their flares. Although soft X-ray observations are often
fit by a single thermal spectrum, there are good reasons to expect
that the X-ray spectra of YSOs are more complex. This is certainly the
case for (magnetically active) M-dwarf stars (Schmitt \etal
1990). Some high-quality ROSAT spectra of YSOs also require a
two-temperature fit (e.g., Skinner \& Walter 1998). The ASCA YSO
spectra also indicate that there is generally a hard as well as a soft
component. The simulations of the Sun as an X-ray star (Peres \etal
2000), based on modeling of YOHKOH Soft X-Ray Telescope observations,
suggest that the high-energy component seen in low-resolution
astronomical satellite measurements of YSOs is associated with small
and large flares. This situation is quite understandable since hard
X-rays are released by energetic flares generated by reconnection
events, whereas soft X-rays are mainly emitted by coronal gas trapped
by large loops of closed stellar magnetic field lines.  Another
conclusion indicated by the available data is that the younger YSOs
are stronger and harder X-ray emitters than the older ones. Again this
is in accord with theoretical ideas about the effects of accretion on
the magnetospheres of YSOs, which are prone to increased magnetic
activity during epochs of enhanced accretion. Of course many of the
limitations in our information on YSO X-rays will be removed by 
measurements of increased sensitivity and angular resolution now being 
made with the {\it Chandra X-ray Observatory} and XMM. The {\it Chandra}
observations (Garmire \etal 2000; Feigelson \etal 2001, private
communication), already support the conclusion from ASCA (Koyama 1999;
Tsuboi 1999) that there are significant differences between younger
and older YSOs with respect to the hardness of spectra and the
frequency of flares.

	In modeling the X-ray spectra of YSOs, we assume that a
soft and a hard X-ray component are present with temperatures $k\TX
\approx 1$\,keV and $k\TX \approx 2-5$\,keV, respectively. Since we
focus on the more active YSOs that produce optical jets (and
correspond to infrared classes I and II), we assume that the
luminosities of the two components are the same order of magnitude,
and that the total X-ray luminosity is substantially larger than the
typical revealed Tauri star, e.g., $\LX \approx 10^{31}$\,erg~s$^{-1}$. Of
course $\LX$ can be even higher during large flares (Feigelson \& Montmerle
1999; Tsuboi 1999; Stelzer \etal 2000).  As in the discussion of the hot spot
radiation in \S 2.3, the calculation of the effects of the X-rays is
complicated by the extended nature of the emitting regions and by the lack of
relevant observational information on their physical properties.  According to
the x-wind model (see Fig.~1 of Shu \etal 1997), soft X-rays can be expected to
arise within the region of closed stellar magnetic field lines (of linear
dimension several $\Rst$) and from the region above the funnel flow and beneath
the helmet dome and kink point (of linear dimension $\Rx$). Hard as
well as soft X-rays may be produced by reconnection events that occur
either in the reconnection ring (in the equatorial plane from 0.75
$\Rx$ to $\Rx$) or along the helmet streamer above the kink point (at
radial distances $\sim \Rx$). We deal with this complicated
geometrical situation by representing the several finite sources by a
set of axial point sources: half of both the soft and the hard X-rays
are assumed to emanate from the origin and the other half from points
displaced along the $z$-axis by $\pm 1.0\Rx$.

	We follow the calculation by GNI of the X-ray ionization rate,
which is based on the energy-smoothed cosmic photoelectric absorption
cross section per H nucleus based on the compilation of Henke, Gullikson, 
and Davis (1993), similar to that used by Morrison and
McCammon (1983),
\be
\label{Xraycross_section} 
\sigma_{\rm pe}(E) = \sigma_{\rm pe}(1\,{\rm keV})({\rm keV}/E)^p;
\ee  
for solar abundances, $p = 2.485$ and $\sigma_{\rm pe}({\rm keV}) =
2.27 \times 10^{-22}$\,{\rm cm}$^2$. We assume that moderate and
high-energy X-rays are most important ($E \geq 1$ keV), and we ignore
the relatively small contributions of the primary and Auger electrons
compared to the dominant secondary electrons. We also assume that the
electron fraction and the temperature are not high enough for the
heavy atoms to be very ionized, so that the main X-rays absorbers are
the K- and L-shells of heavy atoms.  In other words, we approximate
the production rate for primary photoelectrons as if the heavy atoms
are in their ground states using equation~(\ref{Xraycross_section}).  In
the format of equation~(\ref{ioneq}), the ionization rate due to the
secondary electrons is
\be
\label{zeta}
{\cal P}_{\rm X} =  \zeta = \frac{1}{4 \pi r^2}
\int_{E_{0}}^{\infty} \frac{\LX(E)}{E}\sigma_{\rm pe}(E) \,
e^{-\tauX}\,
\left (\frac{E}{\epsilon_{\rm ion}} \right ) dE\,,
\ee
where $\LX(E)$ is the X-ray luminosity per unit energy interval, 
$\epsilon_{\rm ion}$ is the energy to make an ion pair (about 36\,eV
for an unionized hydrogen-helium mixture, according to 
Dalgarno, Yan, \& Liu 1999), and $\tauX$ is the X-ray optical depth.  
\be
\tau_x \equiv \sigma_{\rm pe}(k\TX)N,  \hspace{0.5in} N = \int _0^r n \, dr,
\ee
We introduce a low-energy cutoff $E_0$ because the smoothing
used to obtain the power-law fit, equation~(\ref{Xraycross_section}),
removes the thresholds in the underlying photoelectric cross
sections. The smallest threshold is 0.0136\,keV, but the operative
cutoff may well be larger due to absorption in the source. 
Defining $\xi = E/k\TX$, we choose $\xi_0 = 0.01$, e.g.,  
$E_0 = 0.1$ keV for a thermal X-ray spectrum with $k\TX = 1$\,keV
because. Below this energy, the secondary electrons no longer dominate
the ionization.  We also need the X-ray heating rate,
\be
\label{Xheat}
{\cal G}_{\rm X}  = \frac{1}{4 \pi r^2}
\int_{E_{0}}^{\infty} \frac{\LX(E)}{E}\sigma_{\rm pe}(E) \,
e^{-\tauX}\,
\left ( y_{\rm heat}E \right ) dE\, , 
\ee
where $y_{\rm heat}$ is the fraction of the X-ray energy that heats
the gas.  Both $\epsilon_{\rm ion}$ and $y_{\rm heat}$ are functions
of energy $E$ and electron fraction $x_{\rm e}$, the latter because of
Coulomb scattering between the secondary and ambient electrons. We
exploit the fact that, at energies much larger than characteristic
atomic energies (measured by the ionization potential), $\epsilon_{\rm
ion}$ and $y_{\rm heat}$ are approximately independent of energy.  We
can then extract the asymptotic factor $1/\epsilon_{\rm ion}$ from
equation~(\ref{zeta}) and the asymptotic factor $y_{\rm heat}$ from
equation~(\ref{Xheat}). The result is that the direct X-ray heating, 
in the format of equation~(\ref{heateq}), can be expressed as
\be
\label{approxGX}
{\cal G}_{\rm X} = y_{\rm heat} T_{\rm ion} \zeta,
\ee
where $T_{\rm ion} = \epsilon_{\rm ion}/k$. As discussed in Appendix B 
and \S 2.4.1, there is also an indirect X-ray heating term 
(Eq.~[\ref{indirectX}]) that arises when the thermal 
effects of collisional excitation and de-excitation of the H atom 
are expressed in terms of Lyman-$\alpha$ cooling (Eq.~[\ref{finalheateq}]). 

The dependence of $\epsilon_{\rm ion}$ and $y_{\rm heat}$ on $x_{\rm
e}$ has been studied by many authors, starting with Spitzer \& Scott
(1969) and most recently by Dalgarno, Yan, \& Liu (1999), who give an
extensive set of references to previous work. We have found that the
parameterization of Shull \& Van Steenburg (1983) for atomic H and He
mixtures is useful in calculating the effects of X-rays on the inner 
x-wind with moderate mass-loss rates and not too large electron
fractions.  Their results are confirmed by Dalgarno \etal (1999) (who
also provide the only theory for situations where the hydrogen is
partly or fully molecular). According to Shull \& Van Steenburg, the
energy to make an ion pair can be written,
\be
{1\over \epsilon_{\rm ion}} = 
{y_{\rm H}\over I({\rm H})}+ {y_{\rm He}\over I({\rm He})},
\ee
where
\be
\label{y}
y_{\rm H} = 0.3908\,\left( 1-\xe^{0.4092}\right)^{1.7592}, \qquad
y_{\rm He} = 0.0554\,\left( 1-\xe^{0.4614}\right)^{1.666}, 
\ee
and $I({\rm H})$ and $I({\rm He})$ are the ionization potentials 
of H and He. The heating fraction is
\be
\label{yheat}
y_{\rm heat} = 0.9971\, \left[1-\left(1-\xe^{0.2663}\right)^{1.3163}\right].
\ee
It should be noted that existing theories of electron energy loss, on
which our ionization and heating rates are based, do not hold much beyond 
$\xe = 0.1$. In particular, equation~(\ref{approxGX}) breaks down in 
the limit $\xe \to 1$, as can be seen from the behavior in this limit 
of equations~(\ref{y}) and~(\ref{yheat}). Fortunately, the maximum electron
fractions encountered in the present calculations rarely exceed $\xe = 0.1$.

Putting all of this together for a thermal spectrum with temperature $\TX$, the
ionization rate at distance $r$ from an X-ray sources is, 
\be
\label{zetathermal}
\zeta \approx
\zetax  \left( {\Rx\over r}\right)^2 
        \left( {k\TX\over \epsilon_{\rm ion}}\right)
	I_p(\tau_{\rm x},\xi_0),
\ee
where $\zetax$ is the primary ionization rate at a distance $r=\Rx$.
A useful numerical form for $\zetax$ is 
\be
\label{zetax}
\zetax \equiv {\LX \sigma_{\rm pe}(kT_x)\over 4\pi \Rx^2 k\TX} 
= 1.13\times 10^{-8} {\rm s}^{-1} 
  \left({\LX\over 10^{30}{\rm erg} {\rm s}^{-1}}\right)
  \left({kT_x\over {\rm keV}}\right)^{-(p+1)}
  \left( {10^{12}{\rm cm} \over \Rx}\right)^2.
\ee
The function $I_p(\tau_{\rm x},\xi_0)$ in equation~(\ref{zetathermal})
describes the attenuation of the X-rays. As discussed by GNI, it
decreases more rapidly with optical depth $\tau_{\rm x}$ than a power
law at large optical depth and flattens out to a constant for very
small optical depth. In the application to protoplanetary disks, the
integral was evaluated numerically, but this is infeasible for
detailed modeling of the x-wind. In Appendix C, we obtain an
asymptotic form for $I_p(\tau_{\rm x},\xi_0)$ using the method of
steepest descents (as did Krolik \& Kallmann 1983 and GNI) and then
develop a simple interpolation method to combine the approximations
for small and large optical depths. 

\section{Ambipolar Diffusion Heating}

Long familiar from thermal considerations of interstellar clouds
(Biermann \& Schl\"uter 1950; Scalo 1977; Mouschovias 1978; Lizano \&
Shu 1987), the heating associated with ambipolar diffusion (e.g.,
Mestel \& Spitzer 1956; Spitzer 1978; Shu II 1992) was first applied
to protostellar winds by RGS and to disk winds by Safier
(1993). Because RGS worked with a prescribed spherically-symmetric
wind, they calculated the drag force of the ions on the neutrals from
the wind equation of motion, i.e., as the net force of gravity plus
acceleration. For T-Tauri stars with mass-loss rates in the range
$10^{-8} - 10^{-7}\,M_{\odot}$\,yr$^{-1}$, RGS (Figure 9) obtained
temperatures in the 4,000 --- 5,000\,K range within 10 $\Rst$ but less
than 100\,K beyond $10^{4} \Rst \approx 10^{16}$\,cm (where the
temperature decreases adiabatically as the 4/3 power of the distance).
In this paper, we include ambipolar diffusion heating from first
principles using the dynamical solution of Shang (1998) described in
\S 2.1.

We use an improved approximation for the volumetric rate of ambipolar
diffusion heating because, unlike the situation in interstellar clouds, 
the ionization level in the wind may not be very small:
\be
\label{ADheating}
\Gamma_{\rm AD} = {\rho_{\rm n}|{\bf f}_{\rm L}|^2
\over \gamma \rho_{\rm i}(\rho_{\rm n}+\rho_{\rm i})^2}.
\ee
Here $\rhon$ and $\rhoi$ are the mass densities of the neutrals and
the ions, respectively, $\fL$ is the Lorentz force, 
\be
\label{Lorentz}
\fL = \frac{1}{4\pi}(\nabla \times {\bf B}) \times {\bf B},
\ee
and $\gamma$ is the ion-neutral momentum transfer coefficient. A short
derivation of equation~(\ref{ADheating}) is given in Appendix D, based
on the approximation that the difference in the acceleration (rather
than the velocity) of the neutrals and ions can be ignored.  When
$\rhoi << \rhon$, equation~(\ref{ADheating}) reduces to the usual one
for low-ionization situations, e.g., Eq.~(27.19) of Shu (II 1992). It
has the important property that it does not become singular as $\rhon$
vanishes.  

In Appendix E, we develop an improved formula for $\gamma$
that takes into account the latest calculations and experiments on the
collision of $\hp$ ions with atomic and molecular hydrogen and with
helium, including exchange scattering in $\hp$ + H collisions. 
For the case of no molecular hydrogen and $x_{\rm He}=0.1$,
equation~(\ref{numericalgamma}) yields 
\be
\label{numgamma} 
\gamma = \frac{2.13 \times 10^{14}}{1- 0.714\xe}
\left [\{3.23 + 
41.0 T_{4}^{0.5}(1 + 1.338 \times 10^{-3}\frac{w_5^{2}}{T_4})^{0.5}\}
x({\rm H}) + 0. 243 \right ]\\
\,{ {\rm cm}^3 {\rm s}^{-1} {\rm g}^{-1}},
\ee
where $T_4$ is the temperature in units of 10,000\,K and $w_5$ is the
slip speed (${\bf w} \equiv {\bf v_{\rm i}} - {\bf v_{\rm n}}$) in km
s$^{-1}$. For temperatures approaching $10^4$\,K, the new $\gamma$ is
an order of magnitude larger than the value used by RGS and elsewhere
in the literature. It agrees with Draine's prescription (1980) only for 
cold clouds. 

The slip velocity can be obtained from equation~(\ref{slip}) of 
Appendix D, 
\be
\label{slipagain}
{\bf w} = {{\bf f}_{\rm L}\over \gamma
\rho_{\rm i}(\rho_{\rm n}+\rho_{\rm i})}.
\ee
By eliminating $\gamma$ from the last two equations, we obtain a
quartic equation for $w$, whose solution permits the coupling 
coefficient $\gamma$ to be calculated and then the ambipolar 
diffusion heating to be found from equation~(\ref{ADheating}). The
contribution from He to $\gamma$ in equation~(\ref{numgamma}) (the
last term) is always small.

	A critical factor in the formula for ambipolar diffusion
heating is the square of the Lorentz force (Eq.~[\ref{Lorentz}]). We calculate
the Lorentz force by numerical interpolation on the global x-wind solution
described in \S 2.1, which is itself an interpolation between an interior and
an exterior (asymptotic) solution. Numerical experiments with different
interpolation grids indicate that we have accurately calculated the Lorentz
force for our approximate solution, but we do not have a good estimate of the
error in the solution itself. However, it is reassuring that our calculations
for a model close to that of RGS are consistent with their results.

Figure \ref{fig1} shows the results for what we will call our fiducial model,
defined in Table~2, except that the parameter $\alpha$ for mechanical
heating (to be discussed in \S 5) has been set equal to zero. In other
words, Figure \ref{fig1} includes all of the processes listed in Table~\ref{tbl-1} except
for mechanical heating. The upper panels show temperature contours and
the lower panels ionization contours in the $\varpi$-$z$ plane. The
spatial dimensions are AU and the scale of the plots increases from
right to left. On the smallest scale (at the extreme right), the range
in $\varpi$ is 10 AU and the range in $z$ is 100\,AU. The temperature
very close to the axis approaches 9,000\,K, but the region hot enough
to excite the forbidden lines, roughly $T > 6,000-7,000$\,K, occupies
only a thin inner layer of the jet with a thickness less than
1\,AU. The extreme right panel corresponds to an angular scale of less
than $0.1''$ for objects at a distance of 150\,pc, one not yet
generally available to direct observation.  The scale has begun to be
explored by recent measurements with the Hubble Space Telescope
(Bacciotti \etal 2000) and with an adaptive optics system on the
Canada-Hawaii-France Telescope (Dougados, Cabrit, Lavalley, and
M\'enard 2000).  On the commonly observed scales shown in the
remaining panels of Figure \ref{fig1}, ranging from 1/3 to 3 arcsec, the wind
is warm at best. The ionization fractions range from a few to 10\%.

\placefigure{fig1} 

An analysis of the terms contributing to the basic ionization and heat
equations, (\ref{ioneq}) and (\ref{heateq}), reveals that a variety of
processes contribute, especially close to the star. For example, the
gas starting out on the last streamlines closest to the axis is ionized
by both X-rays and by photoionization of the $n=2$ level and the
negative ion of the H atom (by stellar and hot-spot radiation). The
total ionization and recombination rates balance approximately at
first, but, within a few AU, the ionization falls below the
recombination rate and the electron fraction decreases slowly with
distance (see section \S 6). For streamlines that start out at larger
angles with respect to the axis, the X-rays are more important than
stellar radiation for ionization.  Because the density decreases more
rapidly with distance along these streamlines, the ionization is
essentially frozen into the streamline and decreases even more slowly
with increasing distance.

Without mechanical heating, X-rays are generally the most important heating mechanism and
adiabatic cooling the most important cooling mechanism. However,
Lyman-$\alpha$ cooling is larger than adiabatic within the inner
10\,AU, and ambipolar diffusion heating competes with X-rays at large
distances beyond 500-1,000\,AU. Because the thermal time scale is much
shorter than the ionization time scale, the general dominance of
adiabatic cooling over X-ray and ambipolar heating means that the
temperature decreases rapidly on every streamline.  Of course, as
already noted, these heating mechanisms are relatively weak.  It is
perhaps not surprising that X-ray heating is not that effective,
because only a small fraction of the system luminosity is in X-rays.
In accord with equation~(\ref{ADheating}), ambipolar diffusion heating
is weak because the X-rays produce a relatively high level of
ionization and because the coefficient $\gamma$ is an order of
magnitude larger than used by previous workers. Much higher
temperatures, along the lines obtained by Safier (1993) for disk
winds, could be achieved without X-rays by using the conventional
small (but incorrect) $\gamma$. For our objective of obtaining
physical conditions that are compatible with the optical observations
of jets, a wind thermal model based on stellar radiation, X-rays, and
ambipolar diffusion heating is clearly inadequate.

\section{Mechanical Heating}

	The previous section makes clear that neither heating by
ambipolar diffusion nor heating by X-rays and the other radiation
fields in the problem suffice to explain the observed emission lines
of the abundant heavy elements in YSO winds and jets.  We therefore
consider whether a small fraction of the macroscopic flow energy in
the x-wind can be tapped as a volumetric heat source.  The physical
basis for this idea resides in the observation that the actual flows
are time-dependent (e.g., in the form of pulsed jets as discussed by
Raga \etal 1990; Raga and Kofman 1992), with the time dependence
generating shock waves or turbulent dissipation when fast fluid
elements catch up with slower ones~\footnote{Hydrodynamic turbulence
has often been invoked for heating interstellar clouds (e.g., Black
1987).  Rodriguez-Fernandez \etal (2001) have recently offered it as
an explanation of the warm clouds observed near the Galactic
Center. McKee and Zweibel (1995) and Ostriker, Gammie, and Stone
(1999) discuss the similarities and differences between the
dissipation of hydrodynamic and MHD turbulence.}.  We will not attempt
a detailed discussion of the physics underlying the
transformation of the fluctuating kinetic energy into heat (which
might profitably be studied by 3-d numerical simulations); instead we
appeal to general dimensional reasoning to parameterize the functional
form of the mechanical heating.

The volumetric change in the kinetic energy of the flow
is represented by the following terms in the fluid equations:
\be
\label{advecenergy}
{\partial \over \partial t}\left( {\rho v^2\over 2}\right)
+\nabla \cdot \left[ \left( {\rho v^2\over 2}\right) {\bf v}\right]
\equiv \Gamma_{\rm mech},
\ee
where $\rho$ and ${\bf v}$ are the local gas density and
flow velocity in an inertial frame at rest with respect
to the central star.  Dimensional analysis suggests that
we replace the above expression by
\be
\label{mechheat}
\Gamma_{\rm mech} = \alpha \, \rho \frac{v^3}{s}.
\ee 
where $s$ is the distance the fluid element has traveled
along a streamline to the location of interest in the
wind, and where we have introduced a
phenomenological coefficient $\alpha \ge 0$ to
characterize the magnitude of the mechanical heating.
A choice $\alpha \ll 1$ corresponds to the assumption
that only a small fraction of the kinetic energy contained
in the flow is dissipated into heat via shock waves
and turbulent decay when integrated over the flow
volumes of interest at the characteristic distances $s$ in
the current problem.

The expression (\ref{mechheat}) could be made exact if
we allowed $\alpha$ to be arbitrarily dependent on the
spacetime coordinates $({\bf x},t)$.  In practice,
we shall make the simplifying assumption that $\alpha$
is a global constant, chosen to obtain a reasonable
lighting up of the {\it entire} wind flow.  With $\alpha \ll 1$,
the scaling with $1/s$ in equation~(\ref{mechheat})
could then be justified on the basis of the propagation 
of weak planar shocks where the velocity jump across
the shock varies asymptotically as the inverse square
root of the distance traveled (see \S 95 of
Landau \& Lifshitz 1959), with the energy deposited
into heat (in the fixed frame) varying as the square of
the velocity jump.  In practice, we prefer to regard
equation~(\ref{mechheat}) not as being derived from specific
dissipative processes, but as a generic model equation
whose form satisfies broad physical considerations
and whose utility comes from its simplicity of application
within the context of small fluctuations about some
mean time-steady flow.

From another perspective, when we remember that $v$ tends to $v_{\rm w} \approx
200-300$\, km s$^{-1}$, it is clear that $\alpha$ must be much less than
unity, for otherwise the wind will get too hot. We can roughly approximate
adiabatic cooling by
\be
\label{approxadiabatic}
\frac{3}{2}v (\frac{dT}{ds})_{\rm ad} \approx - \eta \frac{3}{2}\frac{vT}{s},
\ee
where $\eta$ is a parameter that is much less than one for streamlines 
close to the axis and of order unity at large angles 
($\eta = 1/3$ asymptotically, in the latter case).
When this expression is balanced against the mechanical heating equation in 
(\ref{mechheat}), we obtain a rough estimate for the temperature
\be
\label{tempest}
\frac{3}{2}kT \approx (\frac{\alpha}{\eta})m v^2.
\ee
From this result, we see that, not only is the collimated jet ($\eta
\ll 1$) much hotter than the un-collimated wide-angle wind, but
$\alpha$ must be much less than one in order to avoid heating
the wide angle wind to extremely high temperatures.  The full scale
calculations discussed in \S 6 show that $\alpha \sim 10^{-3}$ yields
jet temperatures in the $5,000 - 10,000$\,K range that have been
deduced from observations of forbidden lines. 

\section{Results}
	We now present the results of calculations for a fiducial or 
reference model, defined in Table~2, and for several variations on it. 

\begin{table}[h]
\begin{center}
\begin{tabular}{ll}   \multicolumn{2}{c}{Table 2. Fiducial Model}                
 \\
\hline
\hline
$M_\ast$	&	$0.8\,M_{\odot}$	\\
$R_\ast$	&	$3.0 \,R_{\odot}$	\\
$R_{\rm x}$     &       $4.8 \,\Rst$	\\
$2 \pi / \Omega_{\rm x}$ &       7.5\,d       \\
${\dot M}_{\rm w}$  & $3.2 \times 10^{-8}\,M_{\odot}\,{\rm yr}^{-1}$ \\
${\bar v}_{\rm w}$  & 195 km s$^{-1}$		\\
$L_\ast$	& $2\,L_{\odot}$		\\
$\LX$		& $4\times 10^{31}\,{\rm erg}\,{\rm s}^{-1}$ \\
$\alpha$ 	& $10^{-3}$			\\	
\hline	
\end{tabular}
\end{center}
\end{table}

As remarked earlier, the parameters have been chosen to represent a
solar-mass YSO in a fairly active phase. The numerical value of $\LX$
pertains to two sources, one in and one above the reconnection ring,
each with a soft and a hard component with individual X-ray powers of
$10^{31}\,{\rm erg}\,{\rm s}^{-1}$. The parameter $\alpha$ has been
chosen to have the order of magnitude $10^{-3}$ on the basis of an
approximate solution for the temperature that includes only adiabatic
cooling and mechanical heating and assumes that the density of a
collimated streamline varies inversely with the distance.

\subsection{Temperatures and Ionization Fractions}

Figure \ref{fig2} shows the temperature and ionization profiles (contours of
constant $T$ and $\xe$) on various spatial scales in the same way as
Figure \ref{fig1} (for no mechanical heating). It is immediately clear from a
comparison of the upper panels of Figures \ref{fig1} and \ref{fig2} that mechanical
heating at this level leads to a much warmer wind.  The electron
fraction in the two sets of lower panels are not that different
because X-rays dominate the ionization in both cases.  The
quantitative changes in $\xe$ in going from Figure \ref{fig1} to Figure
\ref{fig2} arise from several temperature-dependent ionization processes:
recombination (decreases with increasing $T$), photodetachment of $\hm$ ($\hm$
abundance increases with $T$), and photoionization of the $n=2$ level of atomic
hydrogen (population increases with $T$). 

The wide range of physical properties manifested in the x-wind imply
that many physical processes are important, although only a few will
dominate at any particular location. For the fiducial model and modest
variations on it, certain processes do play a more global role. For
ionization, they are X-ray ionization and radiative recombination; for
heating, X-rays and mechanical heating; and for cooling,
Lyman-$\alpha$ and adiabatic cooling. We find that the relative
contribution of a process varies from streamline to streamline and
also with distance along an individual streamline. For example,
photoionization of the $n=2$ level of atomic hydrogen is the most
important ionization process within a few AU of the star for the last
10\% of the streamlines close to the jet axis. X-rays then take over
and are effective over distances of 10-20\,AU for producing the
initial ionization of these streamlines. At larger distances, the
total ionization rate is smaller than the recombination rate. Since
the recombination time scale is longer than the dynamical time scale,
the ionization tends to get frozen into the wind and $\xe$ decreases
slowly with increasing distance, as shown in the lower left panels of
Figures \ref{fig1}-\ref{fig3}.  This behavior has been found in several jets of
young stars, e.g., Dougados \etal (2000), Lavalley-Fouquet, Cabrit, and
Dougados (2000), Bacciotti, Eisl\"offel, and Ray (1999), and Bacciotti and
Eisl\"offel (1999). For most of the streamlines, the X-rays are responsible for
setting up the initial ionization of the flow.  Similarly, the inner wind out
to 10\,AU is mainly heated by X-rays, but mechanical heating dominates most of
the rest of the flow. 

In addition to adiabatic and Lyman-$\alpha$ cooling, gas on the inner
streamlines close to the source is cooled by several atomic hydrogen
processes discussed in \S 2.4, mainly recombination and $\hm$ cooling
(equations~[\ref{recombcool}] and equations~[\ref{hminuscool}],
respectively). Within a short distance from the source, adiabatic and
Lyman-$\alpha$ cooling take over and eventually adiabatic cooling
dominates. The transition to adiabatic cooling occurs more rapidly for
the lower streamlines. The cooling by the forbidden lines of O\,I and
S\,II also contribute significantly on the inner, collimated
streamlines, eventually dominating adiabatic for the inner 10\% of the
streamlines and becoming one-third as strong as adiabatic for the
inner 25\% of the streamlines. For most of the rest of the
(uncollimated) wind, forbidden line cooling is unimportant.

\placefigure{fig2} 

Although the wind for the fiducial case (Figure \ref{fig2}) has an ionization
fraction in the right range, it is not hot enough to emit the
forbidden lines at the levels observed in the brightest jets. We can
achieve the desired temperatures by increasing the coefficient
$\alpha$ for mechanical heating. For example, Figure \ref{fig3} shows the
temperature and electron fraction profiles when $\alpha$ is increased
by a factor of two to $2 \times 10^{-3}$.  The temperature is
increased by almost a factor of two, and the electron fraction by
about 20\%. 

\placefigure{fig3} 

\subsection{Synthetic Images}

Figure~4 shows synthetic images of the wind in the forbidden
lines of S\,II $\lambda$6731 (left) and O\,I $\lambda$6300 (right) for
the case illustrated in Figure 3, as they might be observed edge-on
and with near-perfect angular resolution. These images are similar to
those constructed earlier by SSG, but with some unimportant technical
differences in the way that we make the image of the innermost regions
of the jet.  The main physical difference with SSG is that here we
calculate the temperature and electron fraction from an almost
first-principles model, rather than assuming constant values for these
parameters. The abundance of OI is calculated on the basis of a theory
that gives the standard result given by Osterbrock (1989) for H\,II 
regions, i.e., O$^+$ is maintained in chemical equilibrium by very 
fast forward and backward charge-exchange (assuming no H$_2$): 
\be
\frac{x({\rm O}^+)}{x({\rm O})}  = 
\frac{8}{9}\exp^{-227/T} \frac{x({\rm H}^+)}{x({\rm H})} 
\approx \frac{8}{9}\exp^{-227/T} \frac{\xe}{1-\xe}.
\ee
The fact that $\hp$ {\rm +} {\rm O} charge-exchange is fast means that
O$^+$ comes rapidly into equilibrium with the slowly-varying electron
fraction. This situation does not generally hold for other atoms where
charge exchange with $\hp$ is much weaker, especially for sulfur.  We are in
the midst of developing a more general theory of the ionization of the
major ionic carriers of jet forbidden lines. In the interim, we assume
for Figure~4 that all of the sulfur is in S\,II (as in SSG).

\placefigure{fig4}

We obtain the appearance of an optical jet in Figure 4 because the
inner x-wind has a stratified density profile that varies
approximately as the inverse square of $\varpi$ (SSG), the distance to the
jet axis, and because the temperature and electron fraction have the
right values to produce forbidden line emission.  The base of the
model jet in Figure \ref{fig4} has a rounded conical shape suggestive
of HST images (e.g., Eisl\"offel \etal 2000) and a horizontal width of
the order of 25\,AU. The width, which depends somewhat on definition,
is determined by the critical density of the transitions, along with
the temperature and electron fraction.  The emissivities integrated
along the line of sight decline very rapidly with horizontal distance
from a cusp close to the axis.  A careful inspection of the images
near the equatorial plane reveals that O\,I $\lambda$6300 is stronger
than S\,II $\lambda$6731, basically because high-temperature and
high-density regions contribute more to the line-of-sight integral of
the emissivity than the lower temperature and lower density regions
that are more important at high altitudes. Here we are seeing the
result of the competition between the lower critical density of the
S\,II $\lambda$6731 transition and the greater intrinsic strength of
the O\,I $\lambda$6300 transition (and larger O abundance). Images
like Figure \ref{fig4} provide a concrete basis for observational
tests of our thermal-chemical theory of the x-wind jet.

\subsection{Additional Parameter-Space Studies}

In addition to the models shown in Figures \ref{fig2}-\ref{fig4}, we have made
some further exploratory calculations without attempting a systematic
search of the model parameter space, defined largely by mass-loss rate
${\dot M}_{\rm w}$, X-ray luminosity $\LX$, and mechanical heating
strength $\alpha$. For example, we have calculated models with larger
values of $\alpha$. Increasing $\alpha$ from 0.002 to 0.005 increases
the temperature by about 2/3 and the electron fraction by about
1/3. Inside the jet ($\varpi \leq 25$\,AU), $T$ is in the 10,000 -
13,000\,K range and $\xe$ is in the 0.03-0.05 range.  This model may
well have astrophysical applications, and its observational aspects
will differ from those of the $\alpha = 0.002$ model in Figures \ref{fig3} and
4. For example, higher-excitation levels of O\,I and S\,II may be
excited and significant abundances of O\,II and N\,II produced. An
important question is whether a warm model, with $\alpha = 0.005$ or
larger, could produce a level of ionization sufficient to produce the
forbidden lines by just collisional ionization of the H atom {\it without}
X-rays.  Setting $\LX = 0$ in the model with $\alpha = 0.005$, we find
that the electron fraction is reduced by about an order of magnitude
and that the temperature is increased even further, e.g., up to
15,000\,K inside the jet and even higher outside. (The decrease in
$\xe$ makes ambipolar diffusion heating effective close in and reduces
some of the cooling). Obtaining wind ionization levels greater than a
few percent by heating without X-rays doesn't work without going to
temperatures beyond the range indicated by the forbidden line
observations. Some type of external mechanism is required to produce
the degree of ionization inferred from observations, and we have 
shown that stellar X-rays are able to do the job.
   
When we examine the upper part of Figure \ref{fig2}, we see that the
temperature of the wide-angle wind is quite high on the largest scale
shown (the upper left panel), roughly between 3,000-6,000\,K; for
$\alpha = 0.002$ (Figure \ref{fig3}) the range is 6,000-9,000\,K. We
are uncertain of the physical significance of this result because
there are no measurements of the wind in this region (at least until
larger distances are reached where the wind collides with ambient
material). One possibility is that the mechanical heating formula used
so far, equation~(\ref{mechheat}), does not hold for the majority of
the streamlines which are strongly divergent (where shockwaves do not
propagate even approximately according to a planar description).
Using other prescriptions (e.g., changing the density dependence in
equation~(\ref{mechheat}) from a linear to a nonlinear dependence), we
find that the collimated jet can be made warmer and the uncollimated
outer wind colder.

The total X-ray luminosity used in the fiducial model (Table~2), $\LX
= 4\times 10^{31}\,{\rm erg}\,{\rm s}^{-1}$, is several times larger
than the values determined by CHANDRA observations of young solar-mass
YSOs in Orion (Feigelson 2001, private communication).  However, the
key physical parameter is the X-ray ionization parameter, $\zeta /n$,
which determines the imprinting of the ionization at the base of the
wind. Thus the essential model parameter is not $\LX$ alone but
something closer to $\LX/{\dot M}_{\rm w}$, at least for small values
of ${\dot M}_{\rm w}$ where X-ray absorption effects are small or
moderate.  We have run models for $\alpha = 0.002$ (to match Figures \ref{fig3}
and 4) where $\LX$ and ${\dot M}_{\rm w}$ are simultaneously decreased
(and increased) by a factor 3. When $\LX$ and ${\dot M}_{\rm w}$ are
decreased by 3, the temperature is reduced slightly and the ionization
fraction increased somewhat more due to the reduction in X-ray
absorption. The net result is a model which is very similar to the
ones shown in Figures \ref{fig3} and \ref{fig4}.  Increasing $\LX$ and ${\dot
M}_{\rm w}$ may also work, despite the fact that the electron factor is
decreased by a factor of 2 because of the increase in X-ray
absorption, simply because the emissivity of the forbidden lines is
determined by the electron density rather than electron fraction. It
should be recalled that $\LX$ can also be greater than $10^{31}\,{\rm
erg}\,{\rm s}^{-1}$ in flares and that changes in $\LX$ and 
${\dot M}_{\rm w}$ are likely to occur on different time scales.

The warm and ionized conditions found for the outer wind in Figures
\ref{fig1}-\ref{fig3} are in stark contrast to the results of RGS because of
our inclusion of X-ray ionization and mechanical heating in the present
calculations. We have already discussed our uncertainty in applying the heating
model of \S 5 to the uncollimated flow in the absence of compelling
observational information about this part of the wind.  It is quite possible
that the wide-angle wind is not as warm as the above figures
suggest. Furthermore, our thermal-chemical model needs to be extended in this
region to include the thermal and chemical effects of molecules. Although not
mentioned in \S 2, we have made a preliminary study of molecular hydrogen,
mainly to insure that the abundance of $\mh$ is negligible in the inner part of
the wind.  The molecular physics and heavy element chemistry of the wide-angle
part of the wind are of considerable interest in connection with the detection
of H$_2$ jets in young embedded YSOs (e.g., Zinnecker, McCaughrean, and Rayner
1998; Stanke, McCaughrean, and Zinnecker 1998), and we plan to return to this
subject in the near future. 

\subsection{Line Ratios}

	In addition to synthetic images of the forbidden line
emission, like those displayed in Figure \ref{fig4}, we can also examine
particular line ratios that are sensitive to the underlying physical
properties of the wind as calculated in this paper. This approach has
been widely used for H\,II regions and planetary nebulae (Osterbrock
1989), and it has been developed into a diagnostic tool for YSO jets
by Bacciotti and Eisl\"offel (1999). Figure \ref{fig5} is a plot of the
S\,II$\lambda$6716/S\,II$\lambda$6731 line ratio vs.~the
S\,II$\lambda$6731/O\,I$\lambda$ 6300 line ratio based on the same
model as the synthetic images in Figure \ref{fig4}, where the jet is viewed
perpendicular to its axis. The former ratio (the ordinate in the
figure) is diagnostic of $n_{\rm e}$ and the latter (the abscissa) is
sensitive to $T$. The blue dots are the ratios formed from the line
intensities for each pixel in the synthetic image of Figure \ref{fig4}, plotted
one against the other.  The dense concentration of blue points at the
top of the ``blue cliff'' arise from distant wind locations with low
temperature and electron density, whereas the points at the lower left
come from close in where temperature and electron density are high.
The red asterisks are data for HH objects from the compilation of
Raga, B\"ohm, and Cant\'o (1996), and the red circles are data for DG
Tau taken from Lavalley-Fouquet, Cabrit, and Dougados (2000).  The
comparison is meant to be illustrative and should not be taken too
literally.  The blue points have been obtained by ``lighting up''
every pixel of a steady-state jet, as modeled by a specific x-wind
model with mechanical heating according to equation~(\ref{mechheat}),
and viewing the jet from a single direction.  The asterisks are observations
of a diverse set of HH objects viewed at a variety of angles and with
a range of spatial resolution.  Under these circumstances, we should
not expect any more than a general kind of agreement.

\placefigure{fig5}

	The successful demarcation by the theory of the region in
Figure \ref{fig5} where observed YSO-jet line-ratios are found is therefore
extremely satisfying.  It confirms that the temperature and electron
density (integrated along the line of sight) of our thermally and
mechanically self-consistent x-wind model are indeed in the right
range to explain the observations of real sources.  It is significant
that this agreement is obtained by adjusting the one free parameter at
our disposal, $\alpha$, since the others are constrained by
independent observations.  We emphasize that we use the same value of
$\alpha$ in the line-ratio plot of Figure \ref{fig5} as in the image in Figure
\ref{fig4}, $\alpha = 0.002$.  Lavalley-Fouquet \etal (2000) have attempted to
correlate a similar data set with a complex shock model that employs a
continuous distribution of shock velocities for each of 5 values of
pre-shock density, ranging from $10^{2}-10^{6}$\,cm$^{-3}$. The
resulting line-ratio diagram then consists of a set of 5 curves or
branches, one for each pre-shock density. Their analysis does not give
the broad range of conditions observed for HH objects, but it is more
successful with the more uniform set of high spatial-resolution data
for DG Tau. The latter data appear to support a shock interpretation
of the line ratios. 

	It is worth commenting that some of the extreme line-ratios,
represented by the red circles in the lower left part of Figure \ref{fig5},
come from the highest spatial resolution measurements available for DG
Tau obtained with adaptive optics (Lavalley-Fouquet \etal 2000).  It
is no coincidence, we believe, that the high temperature and
electron-density conditions required to produce such ratios occur in
the theoretical model near the base of the observed flow. By
increasing the mass-loss rate slightly, these data would lie closer to
the main body of the theoretical blue points in Figure \ref{fig5}.  The
observations of Lavalley-Fouquet \etal (2000) reinforce the importance
of carrying out spectroscopic measurements at a spatial resolution
sufficient to probe the extreme conditions close to the source of the
jet.  It is also noteworthy that the three red asterisks in the lower
right of the diagram that fall outside the envelope of blue dots are
all associated with strong bow shocks (Raga \etal 1996).  The large
jumps experienced across strong bow shock are evidently not well
represented by the simple formula, equation~(\ref{mechheat}), 
with a (weak-shock)
value $\alpha = 2\times 10^{-3}$. It should also not be too surprising that
some data points lie outside of the theoretical range in Figure \ref{fig5},
considering that the blue dots have been calculated for a single
viewing angle of one specific jet model, whereas the data sample a
wide range of objects viewed under different conditions.

The main purpose of the line-ratio plot in Figure \ref{fig5} is to ensure that
the physical properties of the x-wind, as calculated in this paper,
provide a sound foundation for quantitative comparisons between theory
and observations.  More detailed comparisons will require tailoring
the theoretical models to the specific parameters of individual
sources and considering additional diagnostics, e.g., the forbidden
lines of other species and the radio continuum emission and radio
recombination lines discussed in the next section. This next level of
modeling would better constrain the adjustable parameters of the
problem as well as test the theory under a wide range of flow and
radiation conditions.  Given the developments of this paper, such
detailed tests are within our grasp, but their implementation
is beyond the scope of the present paper.

\section{Discussion}

	The results presented in \S 6 confirm that the
thermal-chemical program described in this paper can provide the basis
for making comparisons between x-wind theory and observations. Using
the forbidden lines S\,II ($\lambda$6731) and O\,I ($\lambda$6300)
from jets as the main example, we have shown that the collimated
portion of the x-wind, when ionized by X-rays and heated mechanically,
emits these lines in a manner strongly suggestive of the actual images
made by observers (Figures \ref{fig4}).  The model also appears capable of
reproducing the measured line ratios (Figure \ref{fig5}).  More definitive
conclusions will require detailed quantitative comparisons between the
theory and observations of individual objects.  Thus we are in the
midst of a study that considers further aspects of the forbidden
lines, such as diagnostic line ratios of additional atoms and ions.
The analysis of forbidden line images obtained at high spatial and
spectral resolution have the potential to provide strong tests of the
predictions that we are now able to make for the x-wind. In order to
explore the jet structure predicted for scales smaller than 25\,AU,
observations with an angular resolution significantly better than
$0.15''$ are required for sources at a distance of 150\,pc. Recent
measurements with the Hubble Space Telescope (Bacciotti \etal 2000)
and with an adaptive optics system on the Canada-Hawaii-France
Telescope (Dougados \etal 2000) have begun to probe jets on this scale.

An independent test of the x-wind model can be made by interferometric
measurements of the radio continuum emission from the hot
partially-ionized gas in the inner wind. Because of the absence of
extinction, radio observations can probe deep into the inner wind
close to the source of the outflow. Thermal radio emission has already
been detected at the centers of more than 100 YSO outflows
(Eisl\"offel \etal 2000; Rodr\'\i guez 1997), and about one fifth of
these have been mapped at high spatial resolution, including a fair
number that can be modeled to test the x-wind (Rodr\'\i guez
2001, private communication). According to the standard theory of
thermal bremsstrahlung emission from a plasma with a uniform
temperature (e.g., Shu I 1991), the optical depth is
\be
\tau_{\rm ff} \approx 6 \times 10^{-28} E \, {\rm cm}^5 \,
(\lambda / 3 {\rm cm})^2 \, T_4^{-3/2}, 
\ee 
where $T_4$ is the temperature in $10^4$\,K and $E$ is the emission
measure.  In the approximation of Paper V, where the density at a
given height above the midplane varies with horizontal distance as
$\varpi^{-2}$, $\tau_{\rm ff} \propto \lambda^2 \, \varpi^{-3}$, and
the size of emission contours of a given intensity level should scale as
$\lambda^{2/3}$, as observed in the best studied cases (see the review of
Anglada 1996). Shorter wavelength observations are then favored for
discriminating the wind from the cooler and lower 
density inner disk.  We expect that the thermal bremsstrahlung
emission for models with the fiducial parameters in Table~2
will be mainly optically thin, except very close to the source. We
plan to synthesize the emission for models of the x-wind as it would
be observed by the Very Large Array. If we are successful in
reproducing the observations, we should be able to determine the
mass-loss rates of thermal jets in a more realistic way than is
currently done with the simple bi-conical model of Reynolds (1986).

Another way of testing the x-wind model with radio observations is
provided by the mm and sub-mm recombination lines emitted by the hot
plasma near the base of wind.  Important kinematic information can be
obtained by sensitive measurements of line shapes at high spatial
resolution. High spectral resolution is also required to detect the
lines in the presence of strong continuum emission by the disk at mm
and sub-mm wavelengths. The hydrogen recombination lines that lie in
the mm and sub-mm bands for $\alpha$ transitions ($n+1 \rightarrow n$)
occur for principal quantum numbers $n$ in the range 25-45. Welch and
Marr (1987) made the first detection of mm recombination lines in the
ultra-compact H\,II region W3(OH) with the H42$\alpha$ line at
86\,GHz. This discovery was soon followed by other detections in
regions of massive star formation (e.g., Gordon and Walmsley 1990) and
by the discovery of masing transitions in MWC 349 by Martin-Pintado,
Bachiller, Thum, and Walmsley (1989). Ground based and space
observations of MWC 349 show that the masing reaches a broad peak at
$n = 19$ or 300 $\mu$m (Thum \etal 1998).  To the best of our
knowledge, radio recombination lines have not yet been detected in T
Tauri stars. Rough preliminary estimates suggest that the radio
recombination transitions produced in the x-wind will be weakly masing for
lines that fall in the familiar sub-mm windows. We expect that some of
these lines will be detected by new sub-mm instrumentation now under
development such as the Sub Millimeter Array (SMA) and the Atacama
Large Millimeter Array (ALMA).  In order to calculate the emission and
to synthesize images as observed with these new instruments, we need
to supplement the program described in this paper with a full
multi-level population calculation for hydrogen and with appropriate
radiative transfer. These developments are now in progress, and we
hope to report soon on the diagnostic prospects of the radio
recombination lines for testing the validity of the x-wind model.

\section{Conclusion}

We have developed a thermal-chemical program that provides the basis
for making detailed predictions for the x-wind model that can be
compared with observations.  The program incorporates new physical
processes, particularly for heating (mechanical) and ionization
(X-rays).  The rate coefficients for all of the underlying microscopic
processes have been re-evaluated and recalculated as required. In some
cases, significantly different values have been obtained from those in
the literature, e.g., the coefficient for ambipolar diffusion heating,
and these should be useful in other problems.

In principle the program can describe a wide variety of flows, all
within the context of the x-wind model. The key astrophysical model
parameters are the mass-loss rate (${\dot M}_{\rm w}$), the X-ray
luminosity ($\LX$), and the mechanical heating strength ($\alpha$).
The first two can be chosen to represent a particular kind of YSO at
some stage of evolution, but a considerable range in these parameters
is allowed by the observations. They may also be variable on short
time scales, as is the case for the X-ray emission. In contrast, we
regard $\alpha$ as a phenomenological parameter. For
the case of an active but revealed source with an optical jet, the
temperature of such jets, as seen in the forbidden lines of oxygen and
sulfur, indicates that $\alpha \approx 2 \times 10^{-3}$.

It is very likely that the three parameters, ${\dot M}_{\rm w}$,
$\LX$, and $\alpha$ are not all independent of one another.  For
example, we might expect that all three parameters decrease as we
proceed from very young and active YSOs to older and less active
ones. In this paper, we have concentrated on sources with optical jets
to illustrate our approach.  The exploration of other cases should
lead to new opportunities for testing the x-wind model. In this 
context, an interesting question is
what kinds of jets occur at earlier evolutionary stages when the
mass-loss rate is much larger than we have used here, $ \sim 3\times
10^{-8}\,M_{\odot}\,{\rm yr}^{-1}$. To answer this question, we are
planning to extend the underlying physics of our model to include the
essential molecular processes that are expected to occur when both the
X-rays and the stellar radiation are more heavily extincted than in
the cases treated in this paper.

The main result of this paper is the demonstration that the x-wind
model, when extended to include thermal and chemical processes, has
the capability to reproduce line ratios as well as images of the
forbidden lines of jets in a self-consistent manner.  We have also
outlined a program for testing the model against the observations of
individual objects, with the goal of better constraining the
parameters of the problem and testing the theory under as wide a range
of conditions as possible.  The developments undertaken in the present
paper now make such detailed tests possible.

\acknowledgments

The authors would be pleased to make available digital versions of the
theoretically calculated emissivities to observers interested in
making detailed comparisons between theory and observation.  This
research has been supported in part by the National Science Foundation
through collaborative research grants to the Berkeley Astronomy and
the NYU Physics departments.  S.L. acknowledges support from
DGAPA/UNAM and CONACyT.  The authors are grateful to Alex Dalgarno,
Dave Schultz, and Pedrag Krsti\'c for advice and help on the cross
sections for the interaction of $\hp$ with H, He, and $\mh$, and they
would like to to thank Luis Rodr\'\i guez for a careful reading of the
manuscript and for his comments.

\appendix

\newpage

\section{Rate Coefficient for $\hm$-$\hp$ Neutralization}

The exothermic channels of the reaction 
\be
\label{hmhpreaction}
{\rm H}^- +  {\rm H}^+ \rightarrow {\rm H}(1) + {\rm H}(n)
\ee
have the energy yields 12.582, 2.648, 0.758, and 0.096 eV for
$n=1,2,3$ and 4.  But curve-crossing considerations and detailed
theoretical calculations (Fussen \& Kubach 1986) indicate that, at
mean center of mass energies less than 2 eV, reactions to the $n=3$
level dominate by a large margin over those to $n=2$. 

The total neutralization cross section has been measured from 0.15-300
eV by Moseley, Aberth, \& Peterson (1970), from 5-2000 eV by Szucs
\etal (1984), and from 30-2000 eV by Peart, Bennett, \& Dolder
(1985). Only the lowest energies below a few eV  are relevant for 
our astrophysical applications, and here the cross sections of Moseley 
\etal decrease with energy $E$ roughly as $E^{-1}$. However, the 
later experiments at higher energies clearly show that the results 
of Moseley \etal are too large by a factor of three. This conclusion is 
supported by the theoretical calculations of Fussen \& Kubach (1986) 
with good potential energy curves.They give an analytic fit to the data
below 3\,eV (after renormalizing the low-energy results), from which 
we obtain the following approximation to the rate coefficient for
$\hm$-$\hp$ neutralization valid below 10,000\,K: 
\be
\label{hmhprateco}
k_{\pm} = 
\left [2.40 \times 10^{-6} T^{-1/2}+ 
       4.96 \times 10^{-9}         + 
       6.46 \times 10^{-11}T^{1/2} +
       7.46 \times 10^{-14}T \right ]\, {\rm cm}^3 {\rm s}^{-1}
\ee
A typical value at $T=10,000$\,K is $3.62 \times 10^{-8}$cm$^3$s$^{-1}$,
which can be compared to the RGS value, $10^{-8}$cm$^3$s$^{-1}$.

\section{Level Population of the Model Hydrogen Atom}

We treat here the effects of collisional and X-ray ionization and
excitation on the population of our model two-level ($n=1,2$) plus
continuum ($c$) H atom. Table~B lists the relevant processes and
associated rates.

\begin{center}
\begin{tabular}{ll}    
\multicolumn{2}{c}{Table B. H Atom Processes}                 \\
\hline
Collisional excitation and de-excitation &$C_{12}, C_{21}$	\\
Spontaneous decay			& $A_{21}\beta_{21}$		\\
Photoionization by the Balmer continuum	& $g_{2c}$		\\
Radiative recombination 		& $\kS(\geq 2)$		\\
Collisional ionization from $n=1,2$	& $C_{1c}$, $C_{2c}$	\\
X-Ray ionization from $n=1,2$		& $\zeta_{1c}$, $\zeta_{2c}$	\\
X-Ray excitation  ($n=1$ to $n \geq 1$)	& $\zeta_{12}$		\\	
\hline
\end{tabular}
\end{center}

The collisional rate coefficients have the form $C= \xe n k(T)$
because we assume that electrons are the most important collision
partners.  We have used Voronov's (1997) fit for $k_{1c}$; at
$T=10^4$\,K, it agrees to within 10\% percent with the simpler formula
used by RGS.  For $k_{2c}$, we fit the rate coefficients given by
Janev \etal (1987) to obtain
\be k_{2c} = 7.37 \times 10^{-10}T^{1/2}e^{-39,471{\rm K}/T}\, {\rm
cm}^{3}\,{\rm s}^{-1}.  \ee For $T=10^4$\,K, $k_{1c}$ is about
$5\times 10^{-8}$ smaller than $k_{2c}$ so that, if the $n=2$ level is
thermally populated, collisional ionizations from the $n=2$ level
proceed at a faster rate than from the ground level. We use the same
collisional de-excitation rate coefficient for the $n=2-1$ transition
as RGS, based on Vernazza, Avrett, and Loeser (1981).
The photoionization rate is given above in \S 2.4.1: 
$g_{2c} = \Wst \, \gAst(2) + \Wh \, \gAh(2)$ (Eq.~[\ref{gA2}]), 
as is the recombination rate, $\kS(\geq 2)$ (Eq.~[\ref{recombrate}]). 
   
The X-ray rates in Table~B are related to the rate $\zeta$ at which
ion pairs are produced, discussed in \S 3. Because the cosmic X-ray
absorption cross section in equation~(\ref{Xraycross_section}) is
normalized to the abundance of hydrogen nuclei, the ionization rate
per unit volume is $\zeta n$, where $n$ is the density of total
hydrogen ($n_{\rm H}$), whereas the rates in Table~B are defined in
terms of level densities. In a H/He mixture, 88\% of the X-ray
produced ions are H$^+$ ions (Dalgarno, Yan, \& Liu 1999).  If we
ignore the small contribution from direct X-ray ionization of atoms in
the $n=2$ level, then $0.88 \zeta n = \zeta_{1c} n_1$. Using the
approximate conservation relations, $n = n_1 + n_c$ and $n_{\rm e }=
n_c$, leads to $0.88 \zeta n = \zeta_{1c} n(1 - x_{\rm e})$, or
\begin{equation}
\label{zeta1}
\zeta_{1c} = \frac{0.88}{1 - x_{\rm e}} \zeta.
\end{equation}
Furthermore, for every H$^+$ ion produced, 1.73 excited H atoms are produced 
(0.37 with $n > 2$ (Dalgarno \etal 1999). Taking into account the 
radiative branching of these excited levels to $n=2$, the rate 
$\zeta_{12}$ is given by
\begin{equation}
\label{zeta12}
\zeta_{12} = \frac{1.34}{1 - x_{\rm e}} \zeta.
\end{equation}

The rate equations based on the processes in Table~B are:
\begin{equation}
\label{n2}
(\zeta_{12} +C_{12})n_1 + x_{\rm e}n \kS(\geq 2) n_c = 
( C_{21} + A_{21}\beta_{21} + g_{2c} + C_{2c} +  \zeta_{2c})n_2  
\end{equation}
\begin{equation}
\label{nc}
x_{\rm e}n \kS(\geq 2)n_c = (C_{1c} + \zeta_{1c})  n_1 
+ (g_{2c}  + C_{2c} + \zeta_{2c}) n_2, 
\end{equation}
\begin{equation}
\label{conserve}
n_1 + n_2 + n_c = n, \hspace{1.0in} n_c = x_{\rm e}n.
\end{equation}

The ratio of the populations in the $n=1,2$ levels can be obtained 
by subtracting equation~(\ref{nc}) from equation~(\ref{n2}):
\begin{equation}
\label{subtracteqs}
(C_{12} + C_{1c} + \zeta_{1c} + \zeta_{12})n_1 = 
(C_{21} + A_{21} \beta_{21}) n_2, 
\end{equation}
\ie, 
\be
\label{n1overn2}
\frac{n_1}{n_2}  \equiv Q =  
\frac{k_{21}n\xe + A_{21}\beta_{21}}{k_{12}n\xe + k_{1c}n\xe + \zeta_{1c} + \zeta_{12}}. 
\ee
The population of the $n=2$ level given in equation of \S 2.1.4 is
obtained by ignoring $n_2$ in the conservation relation, 
equation~(\ref{conserve}): $n_1 \approx (1 - \xe)n$. 
A quadratic equation for the steady electron fraction
can also be obtained from equation~(\ref{nc}), but in this work we
find $\xe$ by integrating the rate equation~(\ref{ioneq}). It may also 
be noted that recombination into and ionization processes out of the 
$n=2$ level drop out of the population ratio, $n_{1}/n_{2} = Q$. 

The thermal implications of the above rate equations can be written as, 
\begin{equation}
\label{heat}
\Gamma - \Lambda = 
(n_2C_{21} - n_1C_{12})E_{21} + g_{2c}n_2\Delta E_{2c} - 
x_{\rm e} n \kS(\geq 2) n_c kT - C_{1c} n_1 I -  C_{2c} n_2 \Delta E_{2c}    
+ \zeta_{1c} \Delta \epsilon_1 n_1, 
\end{equation}
where $\Delta \epsilon_1$ is the X-ray heating energy when an 
ion pair is generated from the $n=1$ level, essentially 
the same as the heating energy used in \S 3,
$\Delta \epsilon_{\rm heat} = y_{\rm heat} \Delta \epsilon_{\rm ion}$.
When we use equation~(\ref{zeta1}) for $\zeta_{1c}$, the last term of 
equation~(\ref{heat}) becomes the usual X-ray heating term (and now includes 
He$^+$), i.e., 
\begin{equation}
\Gamma - \Lambda = 
(n_2C_{21} - n_1C_{12})E_{21} + g_{2c}n_2\Delta E_{2c} - 
x_{\rm e} n \kS(\geq 2) n_c kT - C_{1c} n_1 I -C_{2c} n_2 \Delta E_{2c}    
+ \zeta  n \Delta \epsilon_{\rm heat}. 
\end{equation}

We can rearrange this result using population balance 
equation~(\ref{subtracteqs}),
\begin{equation}
(n_2 C_{21}-n_1 C_{12})E_{21} = -A_{21}\beta_{21}n_2 E_{21}
			+ (C_{1c} + \zeta_{1c} + \zeta_{12}) E_{21}n_1, 
\end{equation}
so as to exhibit the conventional Lyman-$\alpha$ cooling:
\begin{eqnarray}
\label{finalheateq}
\Gamma - \Lambda &=& - A_{21}\beta_{21}n_2 E_{21} 
	           + g_{2c}n_2\Delta E_{2c} 
		   - x_{\rm e} n \kS(\geq 2) n_c kT 
		   - \frac{1}{4} (C_{1c} n_1 + C_{2c} n_2) I \nonumber\\
&+& \zeta n  \left (\Delta \epsilon_{\rm heat} + 2.22 E_{21} \right ). 
\end{eqnarray}
We note that this form of the net heating has two X-ray terms, which
we may refer to as direct and indirect heating. The direct term is the
actual X-ray heating, e.g., as calculated by Shull \& Van Steenburg
(1983) and by Dalgarno, Yan, \& Liu (1999), that arises from elastic
collisions of (X-ray generated) electrons with the atoms and electrons
of the partially ionized plasma. The indirect term, proportional to
$E_{21}$, arises when the net thermal effect of collisional excitation
and de-excitation is expressed in a form where the conventional
Ly-$\alpha$ cooling appears. The large probability for inelastic 
scattering by secondary electron implies that the wind has a 
diffuse radiation field which we ignore for simplicity, 
except for trapped Ly-$\alpha$ radiation.
  
\section{X-ray Attenuation}

The integral that appears in equation~(\ref{zetathermal}), $I_p(\tau_x,\xi_0)$, 
describes the attenuation of X-rays for a thermal spectrum 
(Krolik \& Kallmann 1983),
\be
I_p(\tauX,\xi_0) \equiv \int_{\xi_0}^\infty
\xi^{-p} \exp\left[ -\left( \xi+\tauX \xi^{-p}\right)\right]\,
d\xi ,  
\ee
where $\xi \equiv E/k\TX$, $E_0$ is a low-energy cutoff, and the X-ray 
optical depth is given by 
\be
\tauX \equiv \sigma_{\rm pe}(k\TX) N,  \qquad N = \int _0^r n \, dr.
\ee

Following Krolik \& Kallman (1983) and GNI, 
we use the method of  steepest descents to derive the asymptotic formula
\be
\label{asymptoticI}
I_p \approx \left( {2\pi\over p^2\tauX +\xi_m^{p+1}}\right)^{1/2}
\xi_m^{-(p-2)/2}\exp[-(\xi_m+\tauX \xi_m^{-p})] \equiv J_p(\tauX),
\ee
where $\xi_m$ satisfies the algebraic equation
\be
\label{xim}
\xi_m^p(p+\xi_m) = p\tauX,
\ee
which we solve using Newton's method with the initial guess $\xi_m =
(p\tauX)^{1/(p+1)}$.  Derived for large
$(p^2\tauX + \xi_m^{p+1})$, equation~(\ref{asymptoticI}) is accurate as long
as $\xi_m > \xi_0$, i.e., even for $\tauX$ as small as
\be
\tau_0 \equiv {1\over p}\xi_0^p(p+\xi_0) = 3.4\times 10^{-3}
\ee
when $\xi_0 = 0.1$. For $\tauX < \tau_0$, we 
derive the expansion (valid for $\xi_0 \ll 1$) by repeated integration by parts:
\be
\label{expansionI}
I_p = \left\{ {\xi_0\over (p-1)} - {[1-p\tauX \xi_0^{-(p+1)}] \xi_0^2\over
(p-1)(p-2)} + \dots\right\}
{\exp[-(\xi_0+\tauX\xi_0^{-p})]\over \xi_0^p} \equiv
K_p(\tauX,\xi_0).
\ee
>From equations~(\ref{asymptoticI}) and~(\ref{expansionI}), we construct 
the approximate fitting formula for all $\tauX$,
\be
\label{fitI}
I_p(\tauX,\xi_0) = \left( {\tau_0\over \tau_0+\tauX}\right)K_p
(\tauX,\xi_0) + \left({\tauX\over \tau_0 +\tauX}\right)J_p(\tauX).
\ee

\section{Ambipolar Diffusion Heating for Large Ion Fractions}

We analyze ambipolar diffusion heating with a two-fluid  model 
for the ions and neutrals which satisfies the equations of motion
\be
\label{neutrals}
\rho_{\rm n}{\bf a}_{\rm n} = \rho_{\rm n}{\bf g} +{\bf f}_{\rm d},
\ee
\be
\label{ions}
\rho_{\rm i}{\bf a}_{\rm i} = \rho_{\rm i}{\bf g} - {\bf f}_{\rm d} 
+{\bf f}_{\rm L},
\ee
where 
\be
\label{accel}
{\bf a} \equiv ({\bf v}\cdot \nabla){\bf v}
\ee
stands for the steady-flow acceleration for in an inertial frame, 
${\bf g}$ is the gravitational acceleration, 
\be
\label{drag}
{\bf f}_{\rm d} = 
\gamma \rho_{\rm n}\rho_{\rm i} ({\bf v}_{\rm i} -{\bf v}_{\rm n}) 
\ee
is the volumetric drag force of the ions
on the neutrals, and ${\bf f}_{\rm L} = (\nabla \times {\bf B})
f\times {\bf B}/4\pi$ is the volumetric Lorentz force.
Addition of equations~(\ref{neutrals}) and~(\ref{ions}) yields the 
equation of motion for the combined ion-neutral fluid,
\begin{equation}
\label{combined}
\rho {\bf a} = \rho {\bf g} +{\bf f}_{\rm L},
\end{equation}
where $\rho \equiv \rho_{\rm n}+\rho_{\rm i}$ is the total
mass density (without the electrons) and 
${\bf a} \equiv (\rho_{\rm n}{\bf a}_{\rm n} 
+\rho_{\rm i}{\bf a}_{\rm i})/\rho$ is the total acceleration. 

The dynamics of the x-wind can be obtained to sufficient
accuracy by ignoring the difference between ${\bf v}_{\rm n}$ and
${\bf v}_{\rm i}$ when we use equations~({\ref{accel}) and~(\ref{combined}) to
compute the acceleration of the neutrals and the ions kinematically (from the
mean velocity ${\bf v} \equiv (\rho_{\rm n}{\bf v}_{\rm n} +\rho_{\rm i}{\bf
v}_{\rm i})/\rho$) and dynamically (from the total force $\rho {\bf g} + {\bf
f}_{\rm L}$). We cannot ignore ${\bf v}_{\rm i} -{\bf v}_{\rm n}$ in the
heating problem because ambipolar diffusion heating vanishes unless the ion and
neutral velocities differ. However, we can use the equation for the
difference in the accelerations obtained by subtracting $1/\rho_{\rm
i}$ times equation~(\ref{ions}) from $1/\rho_{\rm n}$ times equation~(\ref{neutrals}):
\be
\label{5}
{\bf a}_{\rm n}-{\bf a}_{\rm i} = \left( {1\over \rho_{\rm n}}
+ {1\over \rho_{\rm i}}\right) {\bf f}_{\rm d} -{1\over \rho_{\rm i}}{\bf
f}_{\rm L} .
\ee
If we now ignore the difference in the accelerations, but not in the
velocities, the left-hand side of \ref{5} is zero and this equation
yields
\be
\label{6}
{\bf f}_{\rm d} = \left( {\rho_{\rm n}
\over \rho_{\rm n}+\rho_{\rm i}}\right){\bf f}_{\rm L}.
\ee
For lightly ionized media, $\rho_{\rm i} \ll \rho_{\rm n}$, we obtain 
${\bf f}_{\rm d}= {\bf f}_{\rm L}$. This is the expected and familiar 
result. When the ionization fraction is low, the ions have relatively 
little inertia and quickly reach terminal velocity governed by the 
balance of frictional and Lorentz forces, i.e., equation~(\ref{ions}) has the
approximate solution, ${\bf f}_{\rm d} = {\bf f}_{\rm L}$.  

We now obtain the slip velocity from equation~(\ref{drag}) for the drag 
force, 
\be
\label{slip}
{\bf v}_{\rm i}-{\bf v}_{\rm n} = {{\bf f}_{\rm L}\over \gamma
\rho_{\rm i}(\rho_{\rm n}+\rho_{\rm i})}.
\ee
This generalizes the lightly-ionized expression by replacing $\rho_{\rm n}$ 
in the denominator by $\rho_{\rm n}+\rho_{\rm i}$.  Expression \ref{slip} is
asymmetric in n and i because the Lorentz force acts only on charged
particles.  Thus, the slip velocity can become large if $\rho_{\rm i}$
becomes very small, but not if $\rho_{\rm n}$ becomes very small (when
the sea of ions, moving under both gravity and the Lorentz force, simply
drags the few neutrals that are present along with the rest of the
almost completely ionized plasma). 

The volumetric rate of heat input into the combined fluid by ambipolar 
diffusion is
\be
\label{AD1}
\Gamma_{\rm AD} = {\bf f}_{\rm d}\cdot({\bf v}_{\rm i}-{\bf v}_{\rm n}), 
\ee
and substitution of equations~(\ref{drag}) and~(\ref{slip}) leads to
\be
\label{AD2}
\Gamma_{\rm AD} = {\rho_{\rm n}|{\bf f}_{\rm L}|^2
\over \gamma \rho_{\rm i}(\rho_{\rm n}+\rho_{\rm i})^2}.
\ee
Notice that $\Gamma_{\rm AD} = 0$ (instead of $\infty$) for a
completely ionized plasma when $\rho_{\rm n} = 0$ but $\rho_{\rm
i}\neq 0$.  Thus equation~(\ref{AD2}) is self-limiting when $T$
approaches and exceeds $10^4$ K (and the gas becomes collisionally
ionized. Although the treatment by Safier (1993) differs 
from ours, his $\Gamma_{\rm AD}$ also vanishes when $\rhon=0$.

\section{The Ion-Neutral Coupling Coefficient For Warm HI Regions}

In order to calculate the ambipolar diffusion heating rate from
\ref{AD2}, we need the momentum transfer rate coefficient $\gamma$
(dimensions cm$^3$\,s$^{-1}$\,g$^{-1}$) introduced in equation~(\ref{drag}).
The velocity field for each species consists of a mean velocity ${\bf v}$ plus
a random velocity ${\bf u}$, which we assume is thermal. Each ion-neutral pair
then gives a contribution to $\gamma$ which involves a double thermal average
of the momentum transfer cross section and can be transformed into (Draine
1986): 
\be
\label{gammadef}
{\bf f}_{\rm d} = \gamma \rhon \rhoi\, {\bf w} =
\Sigma_{jk} n_j n_k m_{jk} \, {\bf w}_{jk} K_{jk}
\ee
where $j$ and $k$ label ionic and neutral species, respectively,
${\bf w} \equiv {\bf v_{\rm i}} -  {\bf v_{\rm n}}$,
${\bf w_{jk}} \equiv {\bf v_j} -  {\bf v_k}$,
$m_{jk}$ is the reduced mass, 
\be
\label{Kjk}
K_{jk} = \frac{1}{\sqrt{\pi}} c_{jk}\,
s_{jk}^{-3} \exp{(-s_{jk}^2)} 
\int_{0}^{\infty}  x^2 \exp(-x^2)\,  
[2xs_{jk} \cosh(2xs_{jk}) - \sinh(2xs_{jk})] \,
\sigma_{jk}(xc_{jk})\,dx ,
\ee
is a momentum transfer rate coefficient, $c_{jk} =
\sqrt{2kT/m_{jk}}$, and $s_{jk} = w/c_{jk}$.  We will approximate
all of the ${\bf w}_{jk}$ by the slip speed ${\bf w}$ of the
two-fluid model used in Appendix D and assume that the random
velocities are all Maxwellian at the same kinetic temperature. When
the momentum-transfer cross section varies inversely with the
velocity, the integral reduces to the usual Langevin rate coefficient
$v \sigma(v)$, whereas for a constant cross section it is (Draine 1986)
\be
\label{rateforconstant}
I= \frac{4}{3} \overline{v} \sigma 
\left [1 + \frac{9}{16}(\frac{w}{\overline{v}})^2 \right ]^{1/2}
\ee
where $\overline{v} = (8 kT/\pi m)^{1/2}$ is the familiar mean 
relative velocity of a pair with reduced mass $m$.
Before evaluating the sum in equation~(\ref{gammadef}), we discuss the 
cross sections that actually occur in this problem. 

\subsection{The Cross Sections}

The model described in the main body of the text deals with warm
atomic regions where the dominant ion is $\hp$, in contrast to
molecular clouds where heavy ions such as C$^+$ and a variety of
molecular ions (notably HCO$^+$) are more important.  The main
collision partners of the ions are H and He atoms. Osterbrock (1961) focused 
on the central role of the (induced)  polarization potential,
\be
\label{polarizationpot}
V= -\frac{1}{2} \alpha_{\rm pol} \frac{e^{2}}{r^4},
\ee
in ion-neutral scattering, where $\alpha_{\rm pol}$ is the
polarizability of the neutral; he calculated the rate coefficient for
momentum transfer between a heavy interstellar-cloud ion and the
dominant gas species, H, He, and H$_2$. Equation~(\ref{polarizationpot})
leads to a temperature-independent rate coefficient (Langevin-type,
$\propto \alpha_{\rm pol}^{1/2}$).  Following Draine (1980),
essentially all authors have taken into account the breakdown of
Osterbrock's treatment at high temperatures, due to interactions of
shorter range than $1/r^4$, by approximating their contribution by a
constant cross section. In the case of Na$^+$ + H$_2$ scattering, for
example, Mouschovias \& Paleologou (1981) (see also Draine, Roberge,
\& Dalgarno 1983) estimated this as a geometric cross section, $1.67
\times 10^{-15}$\,cm$^2$.  This guess has not been borne out by the
recent quantum calculations of Flower (2000), where the
para-H$_2$-HCO$^+$ cross section varies as $E^{-1/3}$ for energies
between $10-10^4$\,K, rather than the Langevin $E^{-1/2}$ dependence,
although the numerical value of the rate coefficient at 20\,K is
essentially the Langevin value).

Draine (1980) also realized that H$^+$+H scattering, basic to our
calculations, is strongly affected by charge exchange because of the
identity of the two nuclei.  There are now good experiments of this
reaction which provide a sound basis for the calculation of the
momentum transfer rate coefficient.  The definitive proof that charge
transfer is significant comes from the merged beam experiment by
Newman \etal (1982), which measures both H$^+$ + H and H$^+$ + D
scattering with an ion optics system that can distinguish between
charge exchange and elastic scattering down to energies as low as
0.1\,eV. They obtain excellent agreement with the theory of Hunter \&
Kuriyan (1977), which is fully quantum mechanical and goes down to
$10^{-4}$\,eV and also agrees with the high-energy theory of Dalgarno
\& Yadav (1953) and other high-energy experiments (e.g., Gilbody
1994).

In addition to the definitive work of Hunter \& Kuriyan (1977) for 
$E > 10^{-4}$\,eV, the momentum transfer cross sections have been 
calculated by Hodges \& Breig (1991) in the same energy region,
and by Krsti\'c \& Schultz (1998) for H$^+$ + H,
H$^+$ + He, H$^+$ + H$_{2}$ for energies $E > 0.1$\,eV. 
It is significant that, for none of the basic molecular ions 
(H$_{2}^{+}$, HeH$^+$, H$_{3}^{+}$), does the scattering 
cross section manifest the pure $E^{-1/2}$ dependence of the Langevin
theory. But when (diffraction) oscillations are averaged out, they all
approximate this energy dependence {\it below a certain energy} $E_1$. 
The three cross sections 
behave differently above $E_1$, as seen in Table~E1,  which gives
{\it approximate} power law fits to the momentum transfer cross section
$\sigma_{\rm mt}(E) = \sigma_{\rm mt}(E_1) (E/E_1)^p$.  
The slopes below and above $E_1$ are $p_1$ and $p_2$, respectively. 

\begin{center}
\tablenum{E1}
\begin{tabular}{|l|l|l|l|l|}    
\multicolumn{5}{c}{Table E-1. Momentum Transfer Power Law Fits}\\
\hline
System  	& $E_1$ (eV)	 & $\sigma_{\rm mt}(E_1)$\,$^1$  & $p_1$	& $p_2$ \\           
\hline
\hline
H$^+$ + H	& 0.01	& 165	& -1/2	& -1/8		\\
\hline
H$^+$ + He	& 1.0	& 9.8	& -1/2	& -1		\\
\hline
H$^+$ + H$_2$	& 5.0	& 5.6	& -1/2	& -2		\\
\hline
\multicolumn{5}{c}{1. Cross section units: $10^{-16}$\,cm$^{2}$}	\\
\end{tabular}
\end{center}

The break points $E_1$ are high enough for H$_2$ and He that the
$E^{-1/2}$ fit is sufficient for $T< 10^4$\,K. For $\hp$+H, however, both
parts of the fit should be retained, although the high energy or
exchange scattering dominates for $T > 100$\,K.

\subsection{Calculations of the Rate Coefficient}

When the momentum-transfer cross sections are approximated by high
and low-energy power laws, as in Table~E1, the rate coefficient
in equation~(\ref{Kjk}) consists of two terms,
\be
\label{2alphas}
 K = \alpha_1 + \alpha_2.
\ee
When we integrate over all energies for the first term, it
becomes a constant, Langevin-type, rate coefficient, which we tabulate in
Table~E2. 

\tablenum{E2}
\begin{center}
\begin{tabular}{|l|l|}    
\multicolumn{2}{c}{Table E2. Rate Coefficients$^1$} \\
\hline
System  	& $\alpha_1$		\\           
\hline
\hline
H$^+$ + H	& 3.23			\\
\hline
H$^+$ + He	& 1.52			\\
\hline
H$^+$ + H$_2$	& 2.83			\\
\hline
\multicolumn{2}{l}{1. Units: $10^{-9}$\,cm$^{3}$\, s$^{-1}$}	\\
\end{tabular}
\end{center}

For H$^+$ + H, we calculate $\alpha_2({\rm H})$ by replacing the $E^{-1/8}$ 
dependence in Table~E1 by the (constant) average value for the interval 
$E=0.01-1.0$\,eV ($1.5 \times 10^{-14}$\,cm$^{2}$), integrating over all 
energy, and using Draine's (1986) 
equation~(\ref{rateforconstant}).
The result is
\begin{equation}
\label{alpha2H}
\alpha_2 = 4.10 \times 10^{-8} {\rm cm}^3\,{\rm s}^{-1} \,T_{4}^{0.5}\,
[1 + 1.338 \times 10^{-3}\frac{w_5^{2}}{T_4}]^{1/2},
\end{equation}
where $T_4$ is the temperature in units of 10,000\,K and $w_5$ is the
drift speed in km s$^{-1}$. Notice that, for $T=10^4$\,K,  $\alpha_2(H)$ is 
more than an order of magnitude larger than $\alpha_1(H)$ due to the 
dominance of charge exchange scattering for $E> 0.01$\,eV.  

For the temperature range of interest, $10^2 < T , 10^4$\,K, $\hp$ + H
dominates over $\hp$ + He and $\hp + \mh$ scattering because of
abundance considerations and also because $\hp$ + H scattering is so
much stronger for $E> 0.01$\,eV. The high-energy contribution
$\alpha_2$ can be ignored for $\mh$ and He because the breakpoints
$E_1$ are larger than for H and because the cross
sections decrease more rapidly than the characteristic low-energy
dependence on $1/v$. Thus $\alpha_1$ in Table~E2 gives a good
approximation (actually upper limit) to the momentum-transfer rate
coefficients for $\hp$ + He and $\hp$ + $\mh$ in the temperature region
of interest

On substituting these results into equation~(\ref{gammadef}), we find that 
the ion-neutral coupling coefficient is
\be
\label{lastgamma}
\gamma = \frac{1}{2m_{\rm H}}
\frac{x({\rm H})[\alpha({\rm H})_2 +\alpha({\rm H})_2] + 
(4/3)x({\mh})\alpha(\mh)_1 + 
(8/5)x_{\rm He}\alpha({\rm He})_1}{x_{\rm H}+2x_{\mh}+4x_{\rm He}},
\ee 
or, numerically,
\begin{eqnarray}
\label{numericalgamma} 
\gamma &=& \left (\frac{2.13 \times 10^{14}}{1- 0.714\xe}\right ){{\rm
cm}^3\;{\rm s}^{-1}\;{\rm g}^{-1}}\nonumber\\  
&\times& \left [\{3.23 + 41.0 T_{4}^{0.5}(1 + 1.338 \times
10^{-3}\frac{w_5^{2}}{T_4})^{0.5}\} x({\rm H}) +  2.21 x({\mh}) + 2.43 x_{\rm He} \right ].
\end{eqnarray}

\clearpage

\begin{figure}
\figurenum{1}
\epsscale{0.9}
\plotone{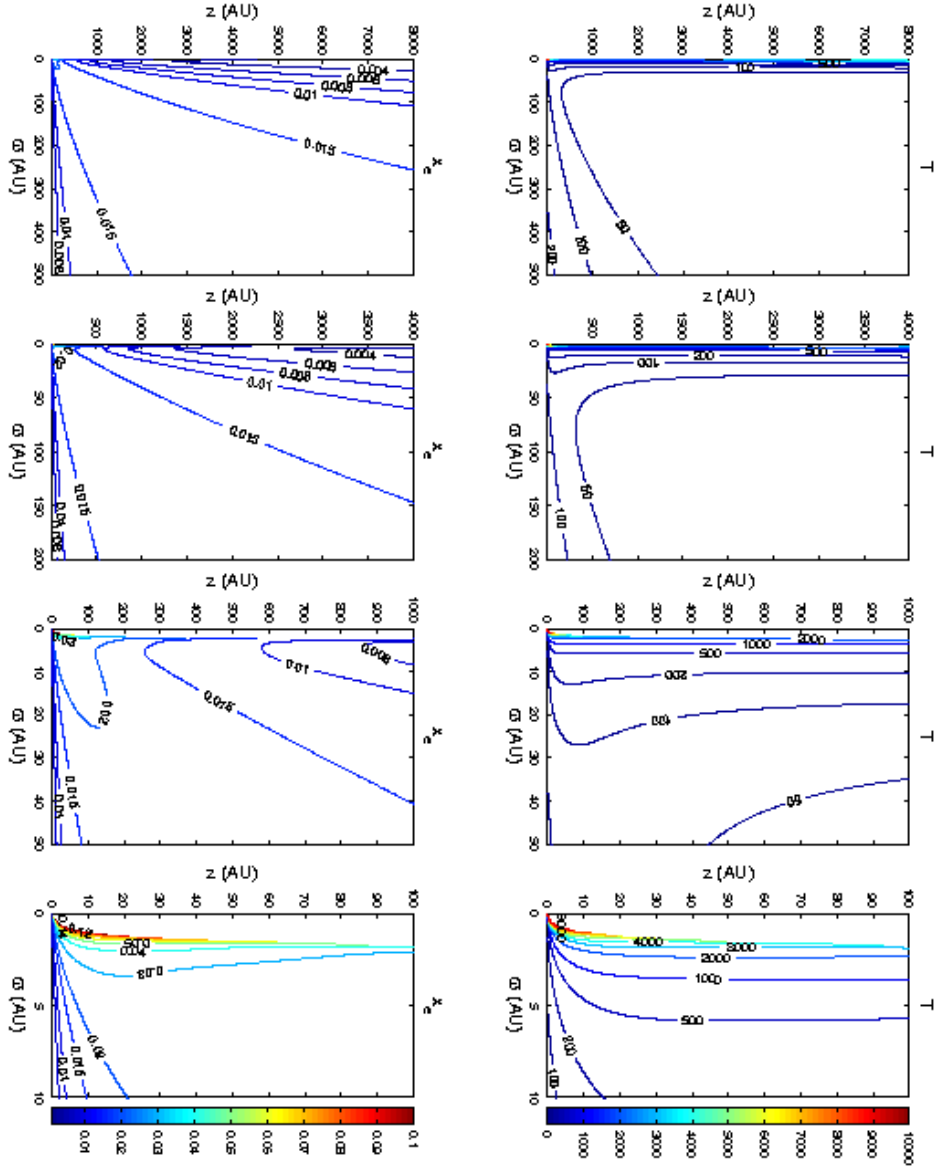}
\caption{Temperature (upper) and ionization (lower) contours in the 
for the $\varpi-z$ plane for the fiducial case (defined in Table 2), 
but with  $\alpha = 0$ (no mechanical heating). The units for the spatial 
scales are AU. Note that the spatial scale proceeds from smaller to larger 
going from right to left.\label{fig1}}
\end{figure}

\begin{figure}
\figurenum{2}
\epsscale{0.9}
\plotone{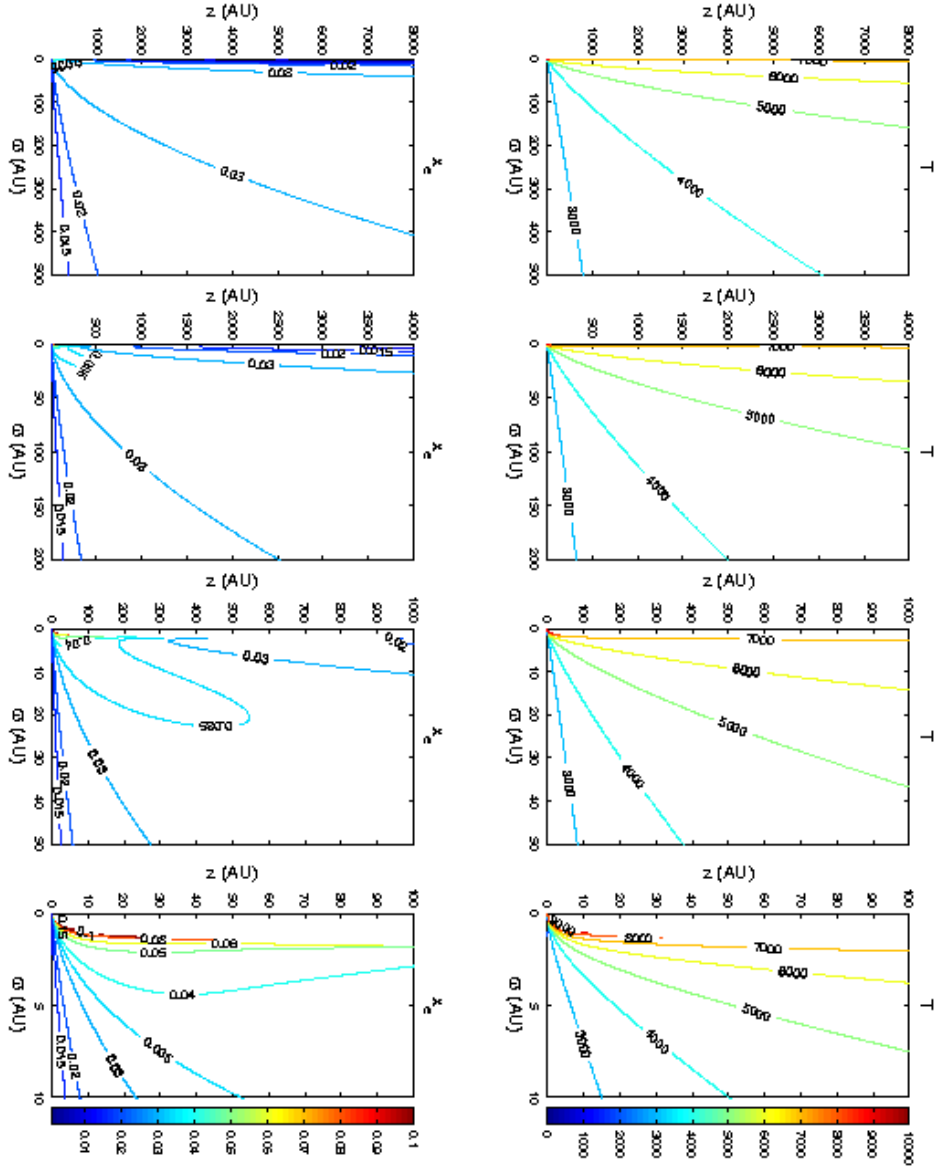}
\caption{Temperature (upper) and ionization (lower) contours 
in the $\varpi-z$ plane for the fiducial case (defined in Table 2). 
The units for the spatial scales are AU.\label{fig2}}
\end{figure}

\begin{figure}
\figurenum{3}
\epsscale{0.9}
\plotone{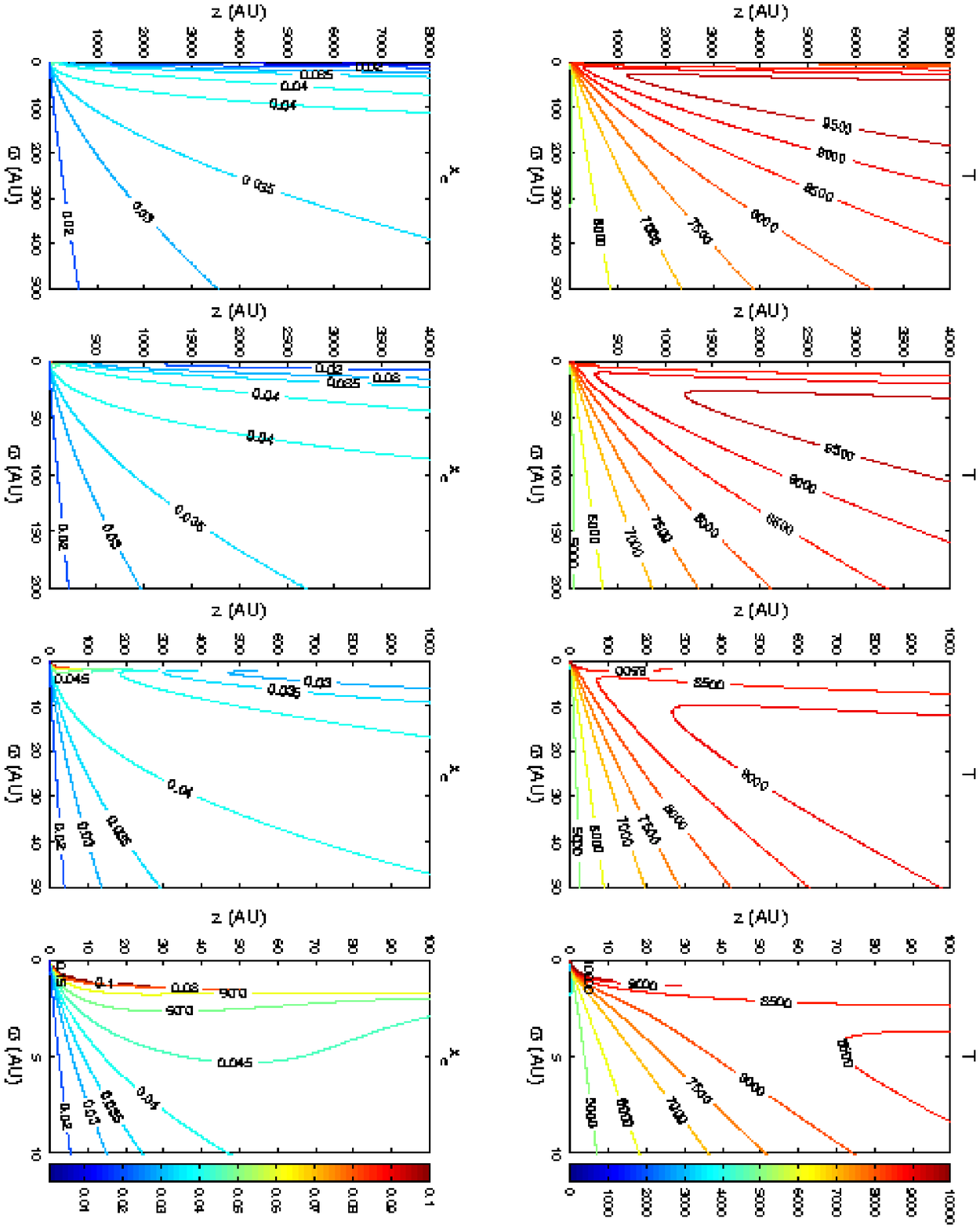}
\caption{Temperature (upper) and ionization (lower) contours 
in the $\varpi-z$ plane for the fiducial case (defined in Table 2),
but with $\alpha = 0.002$. The units for the spatial scales are AU.\label{fig3}}
\end{figure}

\begin{figure}
\figurenum{4}
\epsscale{0.9}
\plotone{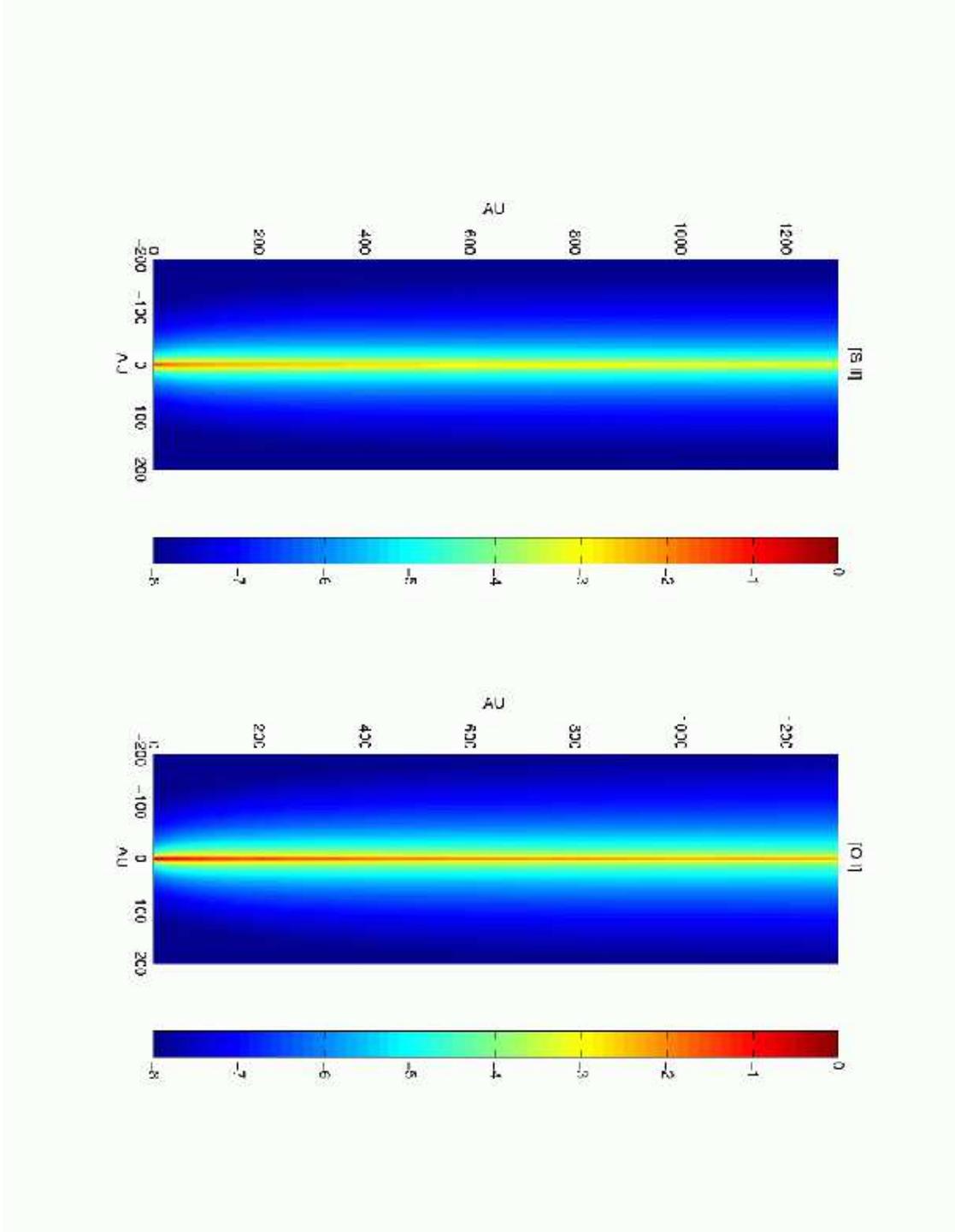}
\caption{Synthetic images of the S\,II $\lambda$6731 (left)
and O\,I $\lambda$6300 (right) brightness for the same model as in
Figure 3 adapting the methods in SSG. The $\log_{10}$ of the integrated
intensity is plotted in units of erg s$^{-1}$ cm$^{-2}$ ster$^{-1}$.\label{fig4}}
\end{figure}

\begin{figure}
\figurenum{5}
\plotone{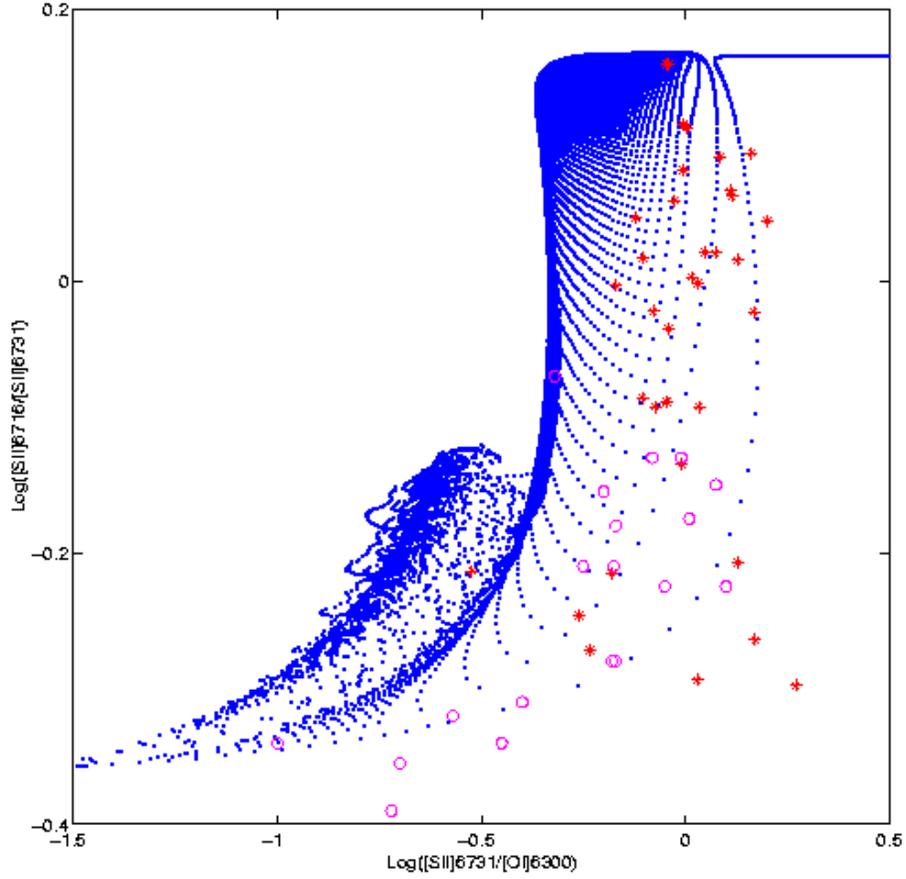}
\caption{S\,II $\lambda$6716/S\,II $\lambda$6731 line ratio
vs. the S\,II $\lambda$6731/O\,I $\lambda$ 6300 based on the synthetic
images in Figure 4, where the jet is viewed perpendicular to its
axis. The model parameters are given in Table 2 but with $\alpha =
0.002$.\label{fig5}}
\end{figure}

\clearpage

\end{document}